\documentstyle[preprint,tighten,aps]{revtex}  
\begin{document}           %
\draft
\preprint{\vbox{\noindent 
Accepted for publication in Physical Review D\hfill hep-ph/9509360\\ 
          \null\hfill MIT-CTP\#2359\\
          \null\hfill INFNCA-TH-94-9}}
\title{Contrasting Real-time Dynamics with Screening Phenomena \\
            at Finite Temperature \\
}
\author{Suzhou Huang$^{(1,2)}$\cite{email} 
        and Marcello Lissia$^{(1,3)}$\cite{email} }
\address{
$^{(1)}$Center for Theoretical Physics, Laboratory for Nuclear Science
and Department of Physics, \\ Massachusetts Institute of Technology,
Cambridge, Massachusetts 02139~\cite{present}\\
$^{(2)}$Department of Physics, FM-15, University of Washington,
Seattle, Washington 98195\\
$^{(3)}$Istituto Nazionale di Fisica Nucleare, 
via Ada Negri 18, I-09127 Cagliari, Italy~\cite{present}\\
and Dipartimento di Fisica dell'Universit\`a di Cagliari, I-09124 Cagliari,
Italy
         }
\date{August 1995; revised version January 1996}
%
\maketitle                 
\begin{abstract}
We discuss the interpretation of Euclidean correlation functions at finite 
temperature ($T$) and their relationship with the corresponding real-time 
Green's functions. The soluble 2+1 dimensional Gross-Neveu model in the 
large-$N$ limit is used throughout as a working example.
First, the real-time bound state, identified as an elementary excitation 
at finite $T$, is solved. The bound state mass, the dispersion relation at 
low momenta, the coupling constant and decay constant are calculated. To 
characterize the structure of the bound state the on-shell form factor is 
carefully introduced and calculated. Then we examine the corresponding 
screening state and contrast the screening mass, coupling constant, decay 
constant and the screening Bethe-Salpeter amplitude with the real-time 
quantities. We find that, although they can be used as qualitative 
indicators in the low-$T$ regime, the screening states at finite $T$ in 
general do not reflect the properties of the corresponding real-time
bound states. Besides, other relevant issues, such as the subtlety of
the real-time manifestation of conservation laws due to some internal
symmetries at $T\ne 0$, the temperature dependence of the pseudoscalar
spectral function and its sum rule, and the high-$T$ limit of the
screening state and its implications to the dimensional reduction,
are also discussed in detail.
\end{abstract}
\narrowtext
\maketitle
\section{Introduction}
\label{intro}

The formulation of QCD, or any other field theory, at finite temperature 
is well-established and several equivalent approaches exist. Nevertheless, 
since finite temperature systems are determined not only by their ground 
states but also by all the excited states, the study of finite temperature 
field theory really involves different physics with respect to the zero 
temperature case and different concepts need to be introduced~\cite{ftrefs}. 

In the existing literature the real-time dynamics is mainly discussed
in a perturbative context. However, we are often interested in 
non-perturbative real-time physics. Typical non-perturbative methods for 
QCD calculation at finite $T$ are the operator product expansion
and lattice simulations. Both these approaches are intrinsically
Euclidean and limited to static properties. The possibility of
extracting real-time information from these Euclidean quantities
needs to be addressed. Due to the lack of experimental data, we do
not have enough intuition to guide our approaches. Therefore, it
is particularly useful to have some exact results which could provide
us with insights into such matter. However, demanding exact solutions 
often implies a sacrifice in the direct phenomenological relevance.

  In this paper, we address these issues in the U(1) chirally symmetric 
Gross-Neveu model~\cite{gn} in $2+1$ dimensions and in the large-$N$ 
limit: this soluble model allows us to perform exact calculations both 
for real-time and screening dynamics. This model is often used in the 
literature~\cite{kocic,gatto} whenever exact results are required and,
besides, some relevance to QCD is also desired. In fact, the Gross-Neveu 
model has several features in common with QCD. For instance, this model
undergoes a chiral phase transition at finite $T$, as it is expected to
happen in QCD, providing a good place for a qualitative modeling of the 
temperature dependence of real-time dynamics. However, some caution is
needed: the Gross-Neveu model is not confining and, therefore, not
every physical interpretation of the singularities associated to
quark and antiquark can be literarily applied to QCD.

  More specifically, we solve the real-time bound state and study its
properties as functions of temperature. We calculate its mass, its
coupling constant, its decay constant and its spectral function.
In addition, we carefully characterize the size of this bound state by 
calculating its on-shell form factor. Similarly, we also solve the 
corresponding screening state by calculating the screening mass, coupling, 
decay constant, and the screening Bethe-Salpeter amplitude. The choice of
screening observables, and of the corresponding real-time ones, has been
suggested from the available lattice calculations~\cite{mscreen,latbs}. 
Eventually, we contrast the real-time observables and the corresponding 
screening ones. 

  While some of the results presented here can be found in the
existing literature, and we include them only to make the
presentation of our points coherent and self-contained, others
are less well-known. Among these new, or insufficiently discussed,
topics we would like to emphasize the following.\\
(1) Due to the lack of Lorentz invariance at finite $T$ a general
amplitude $A(\omega,\bbox{p})$ has often very different functional
dependences on $\omega^2$ and $\bbox{p}^2$. An obvious consequence
of this non-uniformness is, for example, the non-covariant
energy-momentum dispersion relation for elementary excitations. 
A less obvious but more interesting consequence is related to how
conservation laws due to some internal symmetries manifest
themselves in various
physical processes. The Goldberger-Treiman relation and the
effective charge of the pion are used as illustrations.\\
(2) To characterize the structure of a bound state at finite $T$ in
a physical way we introduce the on-shell form factor in the elastic
limit. The peculiarity of the relevant kinematic condition and the
existence of an additional form factor are carefully addressed.
Then we isolate the singularity structure that characterizes
the on-shell form factor as a function of the spatial momentum
transfer and, therefore, determine the size of the bound state.\\
(3) We derive an exact sum rule, which states that the zeroth moment 
of the pseudoscalar spectral function is temperature independent.\\
(4) The screening Bethe-Salpeter amplitudes in scalar and pseudoscalar
channels are calculated explicitly.\\
(5) We obtain the asymptotic formula for the screening mass in the
high-$T$ limit and show that this result demonstrates the failure of
dimensional reduction in this model. The reason of this failure is
also explained.

  Strictly speaking, the 2+1 dimensional Gross-Neveu model with a
continuous symmetry has no finite-$T$ phase transition beyond the
leading order in $1/N$, due to the severe infrared singularity
associated with massless Goldstone bosons. One might question
the relevance of our results at leading order in $1/N$.
This problem can be avoided by working in $2+\epsilon$ spatial
dimensions ($0<\epsilon<1$), where the infrared catastrophe is
absent. Since the limit $\epsilon\rightarrow 0$ is smooth at
the leading order in $1/N$, our results presented in this paper
should be understood as such. We could have also addressed the
same questions in the 3+1 dimensional 
Nambu-Jona-Lasinio model~\cite{hatsuda}. However, since our present
interest is more related to answering conceptual questions rather than
finding direct phenomenological applications, we prefer to use the
2+1 Gross-Neveu model because of its renormalizability. Moreover,
we believe that the qualitative physics associated with the chiral
restoration at finite $T$ is similar in all these models, independently
of their dimensionality. In any case, we shall take notice of those 
results that are directly linked to the specific dimensionality.

The paper is organized as follows. First, we give, in Sec.~\ref{realt},
a brief pedagogical review of how real-time dynamics is formulated at 
finite $T$. Then, in Sec.~\ref{GN}, we introduce the model and collect
all those results that are useful in the rest of the paper. The real-time 
pseudoscalar bound state and its properties are calculated in 
Sec.~\ref{BS}. In this same section we also discuss the 
Goldberger-Treiman relation, the on-shell form factor and derive
a sum rule for the spectral function. The screening state is discussed 
in Sec.~\ref{screen}, where we also contrast minutely the real-time bound 
state and the screening state. The last subsection of Sec.~\ref{screen} is 
dedicated to the high temperature limit of the screening state and to the
implications that this result has for dimensional reduction. We reserve
Sec.~\ref{concl} to a summary of our work and to conclusions.
\section{General Interpretation}
\label{realt}
In this section we review the connection between correlation functions, 
which are the typical output of theoretical calculations, and the 
real-time response of a system to external perturbations.
The following material is most likely well-known to experts~\cite{negele};
nonetheless, since we are specifically concerned with the proper 
interpretation of Euclidean correlation functions, we find useful to 
include it here as a convenient reminder.
 
An experimentalist perturbs a system with an external probe, and then
measures the effect of this perturbation with some detector. In a given
theory we describe this external perturbation using an interaction
Hamiltonian of the form
\begin{equation}
\hat{H}_{\text{ext}}(t) =
\int d\bbox{x}J(\hat{\psi}(t,\bbox{x}))V_{\text{ext}}(t,\bbox{x})\, ,
\quad\quad\text{with}\quad
V_{\text{ext}}(t<0,\bbox{x}) = 0\, ,
\end{equation}
where $V_{\text{ext}}(t,\bbox{x})$ is the external perturbation, which
is switched on at $t=0$, and couples to the system through the current 
$J(\hat{\psi}(t,\bbox{x}))$ (the hat indicates operators). The system
is described by a state, or ensemble of states of the Hamiltonian, and
the measurement of the response to the perturbation is performed through
another current, which can be the same as the one that couples to the
perturbation. For example, we may perturb the system with
an external electric field, which couples to the charge density, and 
measure the resulting change in the same charge density.
Therefore, we typically measure the change of the expectation
value of a given current due to the presence of the perturbation:
\begin{equation}
\delta \langle J(\hat{\psi}(t))\rangle \equiv
 \langle J(\hat{\psi}(t))\rangle_{\text{ext}} 
 - \langle J(\hat{\psi}(t))\rangle_{0} \, ,
\end{equation}
where $\langle \hat{A} \rangle_{\text{ext}}$ is the expectation value 
of the operator $A$ in the state (or ensemble of states) describing
the system in the presence of the external perturbation, and
$\langle A \rangle_{0}$ the corresponding expectation value in the
unperturbed system. Here and in the following we drop the spatial
label for simplicity.

If the perturbation is weak, we can expand in the perturbation,
and keep only the term linear in the external field. Then the linear
response of the system to the perturbation is simply proportional 
to the retarded correlation function:
\begin{mathletters}
\begin{equation}
\label{linres}
\delta \langle J(\hat{\psi}(t))\rangle =
\int_{-\infty}^{\infty} dt' G^R (t-t') V_{\text{ext}}(t'\/) ,
\end{equation}
where the retarded correlation is defined as
\begin{equation}
G^R (t) = \theta(t) \frac{\text{Tr} \left( e^{-\hat{H}_0/T} 
                    [J(\hat{\psi}(t))\, ,\, J(\hat{\psi}(0))]\right)}
                    {\text{Tr} \left( e^{-\hat{H}_0/T}\right)} \,
\end{equation}
and $\hat{\psi}(t) = e^{it\hat{H}_0}\hat{\psi}(0) e^{-it\hat{H}_0}$,
$\hat{H}_0$ is the unperturbed Hamiltonian, and we consider a system
in thermal equilibrium at temperature $T$.
\end{mathletters}
The physical content of this equation is known as the
fluctuation-dissipation theorem, i.e. correlation functions of
a system describe not only the correlations (fluctuations) of
the system in a given channel, but also its linear response
(dissipation) to weak external perturbations.

We say that the system has a (real-time) elementary excitation in the
channel described by a given current, if we get a resonant (very large)
response to the corresponding external perturbation at a given frequency.
If we rewrite Eq.~(\ref{linres}) in frequency-momentum space (restoring
the spatial dependence)
\begin{equation}
\label{freres}
\delta \langle J\rangle (\omega,\bbox{k})= 
\tilde{G}^R (\omega,\bbox{k}) 
\tilde{V}_{\text{ext}}(\omega,\bbox{k}) \, ,
\end{equation}
we can see that the appearance of a pole in the correlator as a
function of $\omega$ gives this kind of resonant response, which will
be called the bound state. It is clear from Eq.~(\ref{freres}) that 
the resonance frequency depends on spatial momentum $\bbox{k}$:
$\omega=\omega(\bbox{k})$. The function $\omega(\bbox{k})$ is the
so-called dispersion relation for the excitation. 

Alternatively, we can use a time-independent external perturbation
($\omega=0$), and study the static response of the system. In this
case, we are interested in the modification of the system in spatial
distribution, e.g. how the effect of the perturbation dies out with
distance (screening). In momentum space there can still exist purely
imaginary poles in the retarded correlator, but now they are
interpreted as pure exponential terms in the decaying response. At
zero temperature only, due to Euclidean invariance for relativistic
theories, there is a one to one correspondence between these poles
in the imaginary momentum and the ones in frequency.

The resonant frequencies of the system (poles of the propagator) are 
important characteristics of the system, but they are not the only
information we get from experiment (and from the propagator). The
strength with which the external probe couples to the system is also
important. For instance, excitations of the  system that do not couple
to the external probe may not be relevant. In the propagator, the
information about the coupling strength is carried by the residue of
the pole at the resonant frequency.

A nice example of the concepts we have just reviewed is the classic
electron plasma probed through the charge density:
$J(t,\bbox{x})=\psi^{\dag}(t,\bbox{x}) \psi(t,\bbox{x})$.
The response of the plasma to a static point charge is just the
well-known Debye screening
\begin{equation} 
\delta \langle J(\bbox{x})\rangle \propto \int dt\, G^R (t,\bbox{x})
        \propto e^{-m_D |\bbox{x}|} \, ,
\end{equation} 
and the corresponding pole in the static propagator,
\begin{equation}
\tilde{G}^R (\omega=0,\bbox{k}\/) 
\propto \frac{1}{\bbox{k}^2 + m_D^2} \, ,
\end{equation}
is the Debye mass: $m_D^2=8\pi n_0 e^2 / kT$ ($n_0$ and $e$ are the
electron density and charge, respectively).

But the plasma also possess real-time excitations, plasmons, with 
characteristic frequency 
$\omega_{\text{ph}}^2(\bbox{k}=0)=4\pi e^2 n_0/m_e$, where
$m_e$ is the electron mass. These excitations are described, near the 
plasmon pole, by the Green's function 
\begin{equation}
 \tilde{G}^R (\omega,\bbox{k}) \propto
 \frac{1}{\omega-\omega_{\text{ph}}(\bbox{k})+i\gamma} \, ,
\end{equation}
where $\gamma$ is the damping rate which gives a finite width
to the plasmon. The frequency-momentum (or dispersion) relation for
plasmons is
$\omega_{\text{ph}}^2 (\bbox{k})=\omega_{\text{ph}}^2(0)+\bbox{k}^2$.

From the preceding example it is quite clear that in general
real-time excitations and screening states describe different
physics. Nevertheless, there is a connection between measurements
in Minkowski and Euclidean space. In general, causality and
unitarity make possible the following dispersion relation for
the correlation function
\begin{equation}
 \tilde{G}^R (\omega,\bbox{k}) =
\frac{1}{\pi}\int_{-\infty}^{\infty} du\,
\frac{\rho(u,\bbox{k})}{u-\omega-i\epsilon}\, ,
\end{equation}
where the spectral function $\rho$ is proportional to the total
cross-section in the given channel. It is important to realize that
the dispersion relation at finite $T$ is in frequency only and the
spatial momentum is treated as a parameter. In the plasmon example,
if the plasmon were the only excitation of the system, the spectral
density $\rho$ would just be the on-shell condition for the plasmon:
\begin{equation}
\rho(u,\bbox{k}) =\lambda^2\,\epsilon(u)\,
\delta ( u^2 - \omega_{\text{ph}}^2 (\bbox{k}) ) \, .
\end{equation}
The power of the dispersion relation is that the physical content
of the correlation is entirely embodied in the spectral function
$\rho$, which determines the correlation in the whole complex
plane, apart from a possible additive polynomial in $\omega$.
For instance, if we analytically continue the correlation 
function to imaginary frequencies, we obtain
\begin{equation}
 \tilde{g}^R (\omega,\bbox{k}) = 
\frac{1}{\pi}\int_{-\infty}^{\infty} du
\frac{\rho(u,\bbox{k})}{u-i\omega}\, .
\end{equation}
This analytically continued correlation function calculated
at the Matsubara frequencies $\omega_n=(2n-1)\pi T$ (for fermions)
is the discrete Fourier transform of the correlator calculated in
the Euclidean region (finite temperature correlator):
\begin{equation}
\label{euccor}
g(\tau) =  \frac{\text{Tr} \left( e^{-\hat{H}_0/T} 
                    J(\hat{\psi}(\tau)) J(\hat{\psi}(0))]\right)}
                    {\text{Tr} \left( e^{-\hat{H}_0/T}\right)} \,
\end{equation}
where
$\hat{\psi}(\tau)=e^{\tau\hat{H}_0}\hat{\psi}(0)e^{-\tau\hat{H}_0}$.
This formula is valid in the interval $0<\tau< 1/T$, while for $\tau$
outside this interval we can use the periodicity condition, which for
fermions reads $g(\tau+1/T)=-g(\tau)$ (it follows from the periodicity
of the trace and the anticommutation property of the fermionic fields). 

Vice versa, if we know the correlation function at the discrete set of 
Euclidean points specified by the Matsubara frequencies, the analytic 
continuation $\omega_n\to -i\omega \pm\epsilon$ provides us with the 
retarded/advanced correlation for continuous real-time frequency. This
continuation is unique only when the expression of the correlation
function is explicitly free of singularity in the entire 
$\omega$-plane except on the real axis and obeys certain convergence
property at $|\omega |\to\infty$~\cite{negele}.

  Finally, due to the lack of Lorentz invariance at finite $T$,
a general Green's function $G(\omega,\bbox{p})$ often has very
different functional dependence on $\omega^2$ and on $\bbox{p}^2$.
For example, $G(0,\bbox{0})$ strongly depends on the order in which
we take the limits $\omega\rightarrow 0$ and $\bbox{p}\rightarrow 0$,
i.e. $G(0^+,\bbox{0})\equiv\lim_{\omega\rightarrow 0}
\lim_{\bbox{p}\rightarrow 0}G(\omega,\bbox{p})$ is usually
different from $G(0,\bbox{0}^+)\equiv\lim_{\bbox{p}\rightarrow 0}
\lim_{\omega\rightarrow 0}G_(\omega,\bbox{p})$.
(The innocent looking function $(x+y)/(x-y)$ in the limit of
$x,\, y\rightarrow 0$ illustrates the relevant subtlety.)
Of course, the physics of the specific process selects which of the
two order is relevant. In particular, $G(0^+,\bbox{0})$ is intrinsically
Minkowskian, because it represents a process whose characteristic time 
is much shorter than the heat-bath response-time. We shall then call this 
kind of processes fast processes. Correspondingly, $G(0,\bbox{0}^+)$ is 
intrinsically Euclidean and the related processes are called slow processes, 
since the thermal environment has enough time to respond to the external 
perturbation. As we will see later, conservation laws due to internal
symmetries only hold in fast processes. In slow processes the apparent
violation of conservation laws should be understood in the sense of the
grand canonical ensemble, not as the violation of the fundamental physics
laws.

\section{The Gross-Neveu Model}
\label{GN}
In this section we introduce the Gross-Neveu model~\cite{gn} 
in $2+1$ dimensions, and present its main features at finite
temperature. Corresponding formulae for the model
in $1+1$ dimensions can be found in Ref.~\cite{huang93}.
\subsection{Lagrangian}
The Gross-Neveu model is defined by the Lagrangian density
\begin{mathletters}
\label{lagr}
\begin{equation}
{\cal L} = \bar{\psi}i{\gamma\cdot\partial}\psi
+{g^2\over 2N}
\left[(\bar{\psi}\psi)^2 + (\bar{\psi}i\gamma_5\psi)^2\right] ,
\end{equation}
where $\psi$ is a 4-component Dirac spinor with the color indices
implicit. Equivalently, we can write
\begin{equation}
{\cal L} = \bar{\psi}i\gamma\cdot\partial\psi
-\bar{\psi}(\sigma+i\pi\gamma_5)\psi
-{N\over 2g^2}(\sigma^2+\pi^2),
\end{equation}
\end{mathletters}
where $\sigma$ and $\pi$ are the auxiliary scalar and
pseudoscalar bosonic fields, respectively. We study this model
in the limit $N\to\infty$ with the coupling constant $g^2$ fixed.
This Lagrangian is invariant under a continuous $U(1)$ chiral
symmetry that is dynamically broken at low $T$. 

\subsection{Effective potential and mass generation}
The standard approach of the effective potential at finite 
temperature~\cite{Jackiw} yields the critical temperature and
the dynamically generated mass.
To the leading order in $1/N$, the effective potential for
$\sigma$ and $\pi$ fields is given by the one-loop expression
\begin{equation}
V_{\text{eff}}(\sigma,\pi)={N\over 2g^2}(\sigma^2+\pi^2)
-2N\int {d^{2}k\over (2\pi)^{2}}\,T\sum_{n=-\infty}^{+\infty}
\ln\left[k^2+\omega_n^2+\sigma^2+\pi^2\right],
\end{equation}
where $\omega_n=(2n-1)\pi T$.
Up to an irrelevant constant
the sum over $n$ can be done yielding
\begin{equation}
V_{\text{eff}}(\sigma,0)={N\over 2g^2}\sigma^2
-2N\int {d^{2}k\over (2\pi)^{2}}
\left\{ \sqrt{k^2+\sigma^2}+{2T}
\ln\left[1+\exp(-\sqrt{k^2+\sigma^2}/T)\right]\right\}\, .
\end{equation}
In this last and the following equations we have used the symmetry 
of the Lagrangian, and rotated the $(\sigma,\pi)$ field in the
$\sigma$ direction making $\pi=0$. Alternatively, we can read
$\sigma$ in the following formulae as representing
$\sqrt{\sigma^2+\pi^2}$. The integral is divergent, and a high
momentum cutoff $\Lambda$ is required. We make $V_{\text{eff}}$
finite by adding a counterterm of the form
\begin{equation}
{\cal L}_{\text{CT}}=-{N\over 2}
(\sigma^2+\pi^2)
\left({\Lambda\over \pi}-{\kappa\over \pi}\right)  \, ,
\end{equation}
i.e. we have defined the bare coupling constant 
$\pi/g^2_B = \pi/g^2-\kappa+\Lambda $, with
$g^2\equiv g^2(\kappa)$ the renormalized coupling constant.

Then, the renormalized effective potential is 
\begin{equation}
\label{pot}
V_{\text{eff}}(\sigma,0)={N\over \pi}
\left\{{1\over3}|\sigma|^3-{\mu\over2}\sigma^2
+{T^3}\int_0^{\sigma^2/T^2}
dz\, \ln\left[1+\exp(-\sqrt{z})\right]\right\} \, ,
\end{equation}
where $\mu\equiv \kappa-\pi/g^2(\kappa)$, which is independent
of the renormalization scale $\kappa$, will turn out
to be the dynamical quark mass at $T=0$. 

The quantum theory is determined by the fluctuations around
the minimum of the effective potential, which we find from
the stationary condition
\begin{equation}
\label{minpot}
{\partial V_{\text{eff}}(\sigma,0)\over\partial\sigma}
\bigg|_{\sigma=\sigma_m}
={N\over\pi}\sigma_m\left\{\sigma_m-\mu+{2T}
\,\ln\left[1+\exp(-\sigma_m/T)\right]\right\}=0\, .
\end{equation}
When $T>T_c$ (high $T$ phase), the expression in braces
is positive definite, and the only solution to Eq.~(\ref{minpot}) 
is $\sigma_m=0$ (symmetric phase).
The critical $T_c$ is the value of $T$ for which the expression
in braces is zero with $\sigma_m=0$, i.e. $T_c=\mu/\ln{4}$.

When $T<T_c$, the expression in braces is zero for the following
value of $\sigma_m$:
\begin{equation}
\label{dynmass}
\sigma_m(T)=\mu\left(1-{T\over T_c}\right)
+{2T}\,\ln\left[1+\sqrt{1-4\exp(-\mu/T)}\right]\, ,
\end{equation}
which is the absolute minimum of the effective potential,
Eq.~(\ref{pot}), while the other stationary point, $\sigma=0$,
is a maximum for these values of $T$. Consequently, the symmetry
is dynamically broken and the corresponding
quark mass is $m(T)=\sigma_m(T)$.

In Fig.~\ref{fig1} we plot the dynamically generated mass as
a function of temperature,
i.e. Eq.~(\ref{dynmass}): it goes exponentially to $\mu$ in
the zero temperature limit, while it approaches zero as 
$2\sqrt{\mu (T_c-T)}$ for $T$ approaching $T_c$ from below.

For later convenience we define the following function of $T$,
which is proportional to the first derivative of
$V_{\text{eff}}(\sigma,\pi)$ with respect to $\sigma^2$ evaluated
at $(\sigma=m,\pi=0)$,
\begin{equation}
R(T)\equiv\mu-m(T)-2T\,\ln[1+\exp(-m(T)/T)]=\left\{
\begin{array}{c} 0 \\ \mu\left(1-{T\over T_c}\right)\end{array}
\quad\begin{array}{l}
\mbox{if $T\le T_c$} \\ \mbox{if $T> T_c$}\end{array}\right.\, .
\end{equation}
\subsection{Pseudoscalar Bubble Graph} 
In this model, the only one-particle irreducible loop graph 
(in the large $N$ limit) that is relevant for our purpose
is the pseudoscalar bubble graph $\Pi_P$. In order to extract the
bound state and screening states, we need this graph in Minkowski
space for general $(\omega,\bbox{p})$. We first use the
imaginary-time formalism to perform the loop integral:
\begin{equation}
\label{bblgr}
i\Pi_P(\omega,\bbox{p}\/)=
NT\sum_{n=-\infty}^{+\infty}
\int^\Lambda\frac{d^2\bbox{k}}{(2\pi)^2}
\text{Tr}\left\{i\gamma_5 \frac{i}{k\cdot\gamma -m} i\gamma_5  
\frac{i}{k\cdot\gamma - p\cdot\gamma-m}
\right\},
\end{equation}
where $k = (i\omega_n,\bbox{k}\/)$, $p = (i\omega,\bbox{p}\/)$, 
$\omega_n=(2n-1)\pi$ with $n=0, \pm 1, \pm 2, \cdots$, and
$\omega=2\pi l$ with $l=0, \pm 1, \pm 2, \cdots$.
The sum over frequencies can be performed by using standard contour
integral techniques~\cite{Jackiw}, and we obtain:
\begin{equation}
\label{psbbl}
i\Pi_P(\omega,\bbox{p}\/)=
-N\int^\Lambda\frac{d^2\bbox{k}}{(2\pi)^2}\frac{\tanh(E_k/2T)}{2E_k}
\left\{
\frac{4k\cdot p}{p^2-2 k\cdot p} - \frac{4k\cdot p}{p^2+2 k\cdot p}
\right\}_{k_0=E_k}\, ,
\end{equation}
where $E^2_k=m^2+\bbox{k}^2$.
The above equation is infinite and can be made finite by
adding the contribution from the counterterm
$i\Pi_{\rm CT}=\partial^2{\cal L}_{\text{CT}}/\partial\sigma^2
              = -N(\Lambda-\pi/g^2-\mu)/\pi$.
We then analytically continue $\omega$ into the entire complex plane.

For calculating the bound state mass, it is sufficient to have
the expression for zero external momentum in the region
$0<\omega^2<4m^2$ for the pseudoscalar channel:
\begin{eqnarray}
i\Pi_P(\omega)&=&{N\over g^2}
+{N\over\pi}\,R(T)
+{N\omega^2\over4\pi}\int_m^\infty dE\, 
\tanh{\frac{E}{2T}}\,{1\over E^2-\omega^2/4} \nonumber \\
&=&{N\over g^2}
+{N\over\pi}\left\{R(T)+{\omega\over4}
\,\ln\left[{2m+\omega\over 2m-\omega}\right]
-{\omega^2\over4}\int_m^\infty dE\, 
{2\over 1+e^{E/T}}\,{1\over E^2-\omega^2/4}\right\}\, .
\label{psbbl2}
\end{eqnarray}

In the region $\omega^2>4m^2$, which is needed for the spectral
function, the bubble develops an imaginary
part, and the corresponding formulae are:
\begin{mathletters}
\label{psbblri}
\begin{eqnarray}
\text{Re}\left[i\Pi_P(\omega)\right] &=& {N\over g^2}
+{N\over\pi}\left\{R(T)+{\omega\over4}
\,\ln\left[{\omega + 2m \over \omega - 2m}\right]
-{\omega^2\over4} P\!\!\!\!\!\int_m^\infty \!\! dE\, 
{2\over 1+e^{E/T}}\,{1\over E^2-\omega^2/4}\right\} \\
\text{Im}\left[i\Pi_P(\omega)\right] &=& 
{N\over 4}\omega\tanh{\frac{\omega}{4T}} \, ,
\end{eqnarray}
\end{mathletters}
where $P\!\!\!\!\int$ indicates the principal part of the integral.

For studying the screening phenomena we need the pseudoscalar
bubble, Eq.~(\ref{bblgr}), in the Euclidean region for $\omega=0$,
i.e. in the static limit. It is convenient to
expand the hyperbolic tangent by using the formula
\begin{equation}
\tanh{x}= \sum^{+\infty}_{n=-\infty}
\frac{x}{x^2+ \left[\pi(n-1/2)\right]^2}\,
\end{equation}
and perform the $\bbox{k}$-integral. The resulting expression is
\begin{equation}
\frac{\pi}{N}i\Pi_P(i\omega\rightarrow 0,\bbox{p})=
{\pi\over g^2}+{R(T)}-T
\sum_{n=-\infty}^\infty\sqrt{\bbox{p}^2\over 4M_n^2+\bbox{p}^2}
\,\ln\left[{\sqrt{4M_n^2+\bbox{p}^2}+\sqrt{\bbox{p}^2}\over
\sqrt{4M_n^2+\bbox{p}^2}-\sqrt{\bbox{p}^2}}\right]\, ,
\label{bubble_p}
\end{equation}
where $M_n^2\equiv\omega_n^2+m^2(T)$.

\subsection{Two-point function}

To the leading order in $1/N$ the retarded two-point function for 
pseudoscalar current involves only a sum over a geometric series of
the pseudoscalar bubble graph:
\begin{equation}
\label{ps2pf}
\langle J_5J_5\rangle^R_T(\omega,\bbox{p})\equiv
\int d\bbox{x}dt\, \exp(i\omega t-i\bbox{p}\cdot\bbox{x})
\theta (t) \langle [J_5(t,\bbox{x}),\bar{J}_5(0,\bbox{0})]\rangle_T =
\frac{i\Pi_P(\omega,\bbox{p})}{1-(ig^2/N)\Pi_P(\omega,\bbox{p})}\, ,
\label{twopoint}
\end{equation}
where
$J_5(t,\bbox{x})=\bar{\psi}(t,\bbox{x})\, i\gamma_5\, \psi(t,\bbox{x})$, 
and we have renormalized the two-point function in such a way that
its lowest order coincides with the renormalized bubble.

When the above equation has a pole in real frequency $\omega$ at
a given spatial momentum $\bbox{p}$, this pole is identified as
the elementary excitation or the bound state. In the static limit 
($\omega=0$), on the other hand, the lowest singularity for 
$-\bbox{p}^2$, denoted by $\tilde{m}^2$, corresponds to the
screening mass. 
\section{Real-time Dynamics}
\label{BS}
  In this section the real-time dynamics is studied. We first solve
the bound state in the pseudoscalar channel, including the bound state
mass, with its dispersion relation for small momenta, the coupling
and decay constants. The Goldberger-Treiman relation is examined
carefully. We then calculate the on-shell form factor and 
the charge radius. Finally, the spectral function and its sum rule 
are derived.
\subsection{Pseudoscalar Bound State}

As discussed in section \ref{realt}, a physical excitation in a
given channel is signaled by a pole in the corresponding correlation
function. Therefore, a bound state in the pseudoscalar channel
(there is no elementary pseudoscalar excitation in the Lagrangian)
exists, if there is a solution to the equation obtained by equating
to zero the denominator of Eq.~(\ref{ps2pf})
\begin{eqnarray}
0&=& 1-{g^2\over N}i\Pi_P(\omega=m_\pi)\nonumber \\
 &=& R(T)+ {m_\pi\over4}
\,\ln\left[{2m+m_\pi\over 2m-m_\pi}\right]
-{m_\pi^2\over 4}\int_m^\infty dE\,
{2\over 1+e^{E/T}}\,{1\over E^2-m_\pi^2/4}\, ,
\label{bseq}
\end{eqnarray}
where we have used the expression for the pseudoscalar bubble, 
Eq.~(\ref{psbbl2}), valid for $0<\omega^2<4m^2$.
This equation has always the solution $m_\pi=0$, which is lower 
than the energy of the unbound particles $2m$, as long as $T<T_c$
($R(T)=0$). The study of the residue associated with this solution
can also help in determining the fate of the bound state when
$T\to T_c$.

As long as we are only interested in the bound state mass,
the bubble calculated at zero momentum $\bbox{p}$ is sufficient.
However, it is also instructive to calculate the energy as a function
of momentum. In fact the system at finite temperature has a preferred
reference frame, the heat bath, and we expect an explicit loss
of covariance for the energy-momentum dispersion relation.
To this end, we perform the calculation of the bound state energy
retaining terms up to the first order in $\bbox{p}^2$ and $\omega^2$
($\omega^2$ is also small in the limit of small $\bbox{p}^2$).
The bubble in this limit can be written as:
\begin{equation}
i\Pi(\omega,\bbox{p})=\frac{N}{g^2}+\frac{N}{4\pi m}
\left\{
\omega^2 \int_1^{\infty}{dx\over x^2} \, \tanh{m x\over 2T}
 -{\bbox{p}^2\over 2} \int_1^\infty dx\,
\left({1\over x^2}+{3\over x^4}\right)\, 
\tanh{m x\over 2T}\right\} \, ,
 \label{full-bubble}
\end{equation}
where we have dropped terms of order 
$\bbox{p}^4$, $\omega^4$, $\bbox{p}^2\omega^2$ and higher. It
should be emphasized that when obtaining Eq.~(\ref{full-bubble})
the bubble is expanded in powers of $\bbox{p}^2$ first, and then
the corresponding coefficients are expanded in powers of $\omega^2$.
Again by solving the bound state equation $i\Pi=N/g^2$, we find
\begin{mathletters}
\begin{equation}
\omega^2(\bbox{p}^2) = v^2(T)\,\bbox{p}^2 + O(\bbox{p}^4 )
\end{equation}
where the function
\begin{equation}
 v^2(T)=\frac{\int_1^{\infty}dx\,(x^{-2}+3x^{-4})\tanh(m x/2T)}
   {2 \int_1^{\infty}dx\,x^{-2} \tanh(m x/2T)}\, ,
\end{equation}
\end{mathletters}
has the following limits: $\lim_{T\to 0}v^2(T)=1$ , as it should,
and $\lim_{T\to T_c}v^2(T)=1/2$. The fact that $v^2(T)<1$ implies
that the speed of the pion is reduced by the thermal environment,
even though the pion is still massless.

As stated in section~\ref{realt}, the bound state coupling
constant to a quark-antiquark pair is given by the residue
of the full pion propagator on the mass-shell
\begin{equation}
\label{g2pi}
g_\pi^2(T)=\left({\partial\over\partial\omega^2}
i\Pi_P(\omega)\right)^{-1}_{\omega=m_\pi}={4\pi m\over N}
\left(\int_1^\infty {dx\over x^2}\,\tanh{m x\over 2T}\right)^{-1}\, ,
\label{g_pi}
\end{equation}
which is also proportional to the residue of the pole of the
correlation function, Eq.~(\ref{ps2pf}), at $\omega=m_{\pi}$.

The dependence of $g_\pi^2(T)$ on temperature is shown in
Fig.~\ref{fig2}. When $T\rightarrow 0$, $m\rightarrow\mu$, and
$g_\pi^2\rightarrow 4\pi\mu/N$ (zero temperature limit). When
$T\rightarrow T_c^{-}$, $m(T) \rightarrow 0$, and
$g_\pi^2(T)\rightarrow 0^{+}$.
In this last case, the coupling $g_\pi^2(T)$ approaches zero
logarithmically: $\lim_{T \to T_c^{-}} g_\pi^2(T)\propto 
(\log{m})^{-1} \propto [\log{(T_c-T)}]^{-1}$. In contrast, the
corresponding coupling in the four dimensional Nambu-Jona-Lasinio
model in the chiral limit approaches a finite constant at $T_c-0^+$,
and then it jumps to zero for $T>T_c$~\cite{hatsuda}.

The bound state solution in the pseudoscalar channel for $T=T_c$
has the same energy of two free quarks, and one might wonder about 
the fate of the bound state in the limit $T \to T_c^{-}$. The
coupling constant result gives us a clear answer suggesting that
the bound state disappears at the phase transition. More precisely, 
we should conclude that the pseudoscalar meson decouples from its
constituents at the phase transition.

\subsection{The Goldberger-Treiman relation}
  We can also introduce a pion decay constant defined as the
residue of the axial current correlation function at the pion
pole (this definition at $T=0$ yields the usual expression
$\langle 0|J_\mu^5|\pi(p)\rangle=-ip_\mu f_\pi$). So we need
to calculate the axial-pseudoscalar bubble graph
$A_\mu(i\omega,\bbox{p})$, defined by replacing in Eq.(\ref{bblgr}) 
one of the $i\gamma_5$ with $\gamma_\mu\gamma_5$.
It is easy to verify that
$A_\mu(i\omega,\bbox{p})={p_\mu} A(i\omega,\bbox{p})$
with
\begin{equation}
A(i\omega,\bbox{p})={NT\over\pi}\sum_n\int_0^1 d\alpha\,
{m\over \omega_n^2+m^2+\alpha\omega(\omega-2\omega_n)
+\alpha(1-\alpha)\bbox{p}^2}\, ,
\label{a_p}
\end{equation}
where $\omega=2\pi l$ is the bosonic Matsubara frequency.
The above equation can not be naively used to analytically
continue $i\omega$ into the entire complex plane, since
$i\omega$ has singularities off the real axis. However, for
calculating the pion decay constant we only need $A$ at
$\bbox{p}=0$. Then the $\alpha$-integral can be done, while
the sum over $n$ can be carried out by the standard contour
integral technique, yielding, 
\begin{equation}
A(i\omega,\bbox{p}=0)={N\over 2\pi}\int_m^\infty dx\,
\tanh\left({x\over 2T}\right)\,{m\over x^2-(i\omega)^2/4}\, ,
\end{equation}
which now can be trivially continued and then put on mass-shell.
Hence, the pion decay constant is given by
\begin{equation}
f_\pi(T)=g_\pi(T)\,A(0^+,\bbox{0})
=g_\pi(T)\,{N\over 2\pi}\,\int_1^\infty {dx\over x^2}
\tanh\left({m x\over 2T}\right)\, ,
\label{fpirt}
\end{equation}
which is valid in the broken phase. Since there is no pion state
in the symmetric phase, $f_\pi(T)$ is not defined when $T>T_c$.
The vanishing of $f_\pi$ at $T=T_c$ merely reflects the fact that
the chiral symmetry is restored. Using the explicit expression
of $g_\pi^2(T)$, the above equation can be rewritten as
\begin{equation}
g_\pi(T)\,f_\pi(T)=2m(T)\, ,
\label{new_gt}
\end{equation}
which is the Goldberger-Treiman relation at finite $T$.
Notice that $g_A=1$ identically in the large-$N$ limit.

If we were not careful
in using $A(0^+,\bbox{0})$ to define $f_\pi(T)$ we would
have got an expression of $f_\pi(T)$ not satisfying the 
Goldberger-Treiman relation. For example, if we put the
axial-pseudoscalar bubble graph on the mass-shell before
taking the limit $\bbox{p}\rightarrow 0$, we would have
effectively used $A(0,\bbox{0}^+)={N\over 2\pi}\tanh(m/2T)$
in Eq.~(\ref{fpirt}). The physical reason that the pion
decay should be regarded as a fast process is that the
decay happens instantaneously and the thermal environment
does not have enough time to respond.

  At this point it is appropriate to comment on the fact that
it is not trivial how, at finite temperature, the conservation
laws that are the consequence of symmetries of a theory manifest
themselves in the real-time dynamics. Ward identities associated
with global symmetries can be straightforwardly generalized from 
the $T=0$ case to finite $T$ in the Euclidean formalism. However, it
is very subtle how to subsequently analytically continue these
relations between Euclidean Green's functions to corresponding
relations between real-time quantities and at which stage one should
put the external lines on their mass-shells. Therefore, one can not 
simply assume that Euclidean Ward identities immediately apply also 
to the on-shell quantities, as was done in Ref.~\cite{shen}. It
appears that the validity of the conservation laws depends on the 
specific mathematical prescriptions, but this dependence is
not arbitrary: it reflects the nature of the formulation
of field theories at finite $T$ in terms of ensemble averages.
Only in fast processes, where the heat bath does not have time
to respond, conservation laws hold explicitly. Whereas in slow
processes, where a measurement always involves the feedback of
the heat bath, conservation laws are no longer manifest. We will
encounter an example of such a ``violation'' in the next subsection.

\subsection{On-shell Form Factor}

  At any temperature the on-shell form factor in the elastic
limit provides a physical measure of the size of an elementary
excitation. In fact, even at finite $T$, we can imagine the
following experiment. In the heat-bath frame, we scatter a
lepton off the pion. Just as in the usual zero temperature
scattering, we select those leptons that have scattered off
an on-shell pion rather than off something else by using the
appropriate kinematic conditions. In our case, this is possible
because the pion is massless, while an unbound quark-antiquark
is massive. The elasticity here is important, since the
kinematics selects out slow processes in which the scattered
lepton sees not only the pion but also the thermal cloud around
it. The size defined this way provides a snap shot of the
elementary excitation immersed in an equilibrated thermal
environment.

The on-shell form factor has also been studied in Ref.~\cite{schulze}.
However, the author apparently did not realize several complications 
that arise when defining on-shell form factors at finite $T$, such as 
the proper on-shell condition at finite $T$ and the appearance of 
additional form factors.

  On general ground, a typical three-point function with two
pion lines and one photon line, as shown in Fig.~\ref{fig4}, has
the following structure in the heat-bath frame
\begin{equation}
f_\mu(p,p')=(p_\mu+p'_\mu)F[p^2,p'^2,p\cdot p']
+\delta_{\mu 0}G[p^2,p'^2,p\cdot p']\, .
\label{f_mu}
\end{equation}
The additional form factor $G[p^2,p'^2,p\cdot p']$ vanishes identically 
at $T=0$ and it is related to the heat bath. To minimize the environmental 
effect and hence to emphasize the intrinsic structure of the pion it is 
natural to use the the invariant function $F$ to define its size; to this
end, we can select the spatial components $f_i(p,p')$, which receive
contribution only from $F$, i.e. we consider a ``magnetic scattering 
process''.

For low energy elastic scattering, the momentum transfer is
$q\equiv p'-p=(0,\bbox{p}'-\bbox{p})$, i.e.
$p_0=p'_0$ and $\bbox{p}^2=\bbox{p}'^2$. Due to the breaking
of the explicit Lorentz invariance at finite temperature, the
on-shell condition for the pion does not imply
$p^2=\omega^2(\bbox{p})-\bbox{p}^2=m_\pi^2$. Therefore, there exist
two independent variables in the on-shell form factors $F$ and $G$,
even in the elastic limit. We choose these two independent variables
to be $\omega(\bbox{p})$ and $\bbox{q}^2=(\bbox{p}'-\bbox{p})^2$.
Then the radius of the pion at finite temperature is defined as
\begin{equation}
\langle r^2 \rangle_T\equiv 
-4\lim_{\omega(\bbox{p})\rightarrow m_\pi}
\Biggl\{\lim_{\bbox{q}^2\rightarrow 0}
{\partial\ln F[\bbox{q}^2,\omega(\bbox{p})]\over\partial\bbox{q}^2}
\Biggr\}\, .
\end{equation}
The factor of 4, rather than the usual 6, is due to the fact that
we are in two spatial dimensions. This definition coincides with
the usual definition at zero temperature. The overall normalization
of $F[\bbox{q}^2,\omega(\bbox{p})]$, which depends on $T$ because
the charge of the pion gets screened
($F[\bbox{q}^2=0,\omega(\bbox{p})]\le 1$ in general),  does not
affect the charge distribution.
The limit of $\omega(\bbox{p})\rightarrow m_\pi$ is to ensure that
the radius is measured in the heat-bath frame. This condition need
not be imposed at zero temperature, due to the Lorentz covariance.
In case one has trouble to imagine a charge radius for a massless
particle, we could have included a small quark mass in the Lagrangian
such that $m_\pi>0$. This formal process is not really necessary
since the limit $m_\pi\rightarrow 0$ is smooth.

  The graph that gives the leading $1/N$ contribution to 
$F[\bbox{q}^2,\omega(\bbox{p})]$ is shown in Fig.~\ref{fig4}.
The advantage of
probing the pion by a photon instead of a scalar, the $\sigma$, is
that meson-pole dominance is avoided, consistent with our intention
to emphasize the intrinsic structure. Using the pion-quark-antiquark
coupling derived in the last section, we have
\begin{equation}
f_i(p,p')=g^2_\pi(T)\, T\sum_{n=-\infty}^\infty
\int{d^2\bbox{k}\over(2\pi)^2}\, {\text{Tr}\gamma_5(k\cdot\gamma+m)
\gamma_5(k\cdot\gamma-p\cdot\gamma+m)\gamma_i
(k\cdot\gamma-p'\cdot\gamma+m)\over
[k^2-m^2][(k-p)^2-m^2][(k-p')^2-m^2]}\, ,
\label{ff-define}
\end{equation}
where $m=m(T)$ is the mass gap. In the elastic limit the momenta are
$k=(i\omega_n,\bbox{k})$, $p=(p_0,\bbox{p})$ and $p'=(p_0,\bbox{p'})$
with $p_0=\omega(\bbox{p})=v(T)|\bbox{p}|+{\cal O}(|\bbox{p}|^3)$.

  Since we are dealing with a slow process, we should enforce the
mass-shell condition at the very beginning. We can take advantage of
the facts that the pion is massless and that the scattering is
elastic and enforce the on-shell condition by setting all external 
Matsubara frequencies to zero. This step is perfectly legal, because 
the Euclidean point $i\omega=i2\pi l$ with $l=0$ happens to lie on the 
real axis and coincides with the point $\omega=0$. Then, the calculation
of the on-shell form factor from Eq.~(\ref{ff-define}) involves the 
following steps: (1) combine the denominators using the Feynman parameter 
representation; (2) carry out the spatial momentum integral; (3) eliminate 
all momentum variables in terms of the two independent variables 
$\bbox{q}^2$ and $\omega(\bbox{p})$; (4) extract 
$F[\bbox{q}^2,\omega(\bbox{p})]$ from $f_i(p,p')$ using Eq.~(\ref{f_mu}); 
(5) take the limit $\omega(\bbox{p})->m_\pi$.
Without going through the details, which are tedious but
straightforward, we directly give the final result
\begin{equation}
F[\bbox{q}^2,\omega(\bbox{p})\rightarrow 0]=
{g^2_\pi(T)\over 2\pi}T\sum_{n=-\infty}^\infty\int_0^1 d\alpha
{1\over \omega_n^2+m^2+\alpha(1-\alpha)\bbox{q}^2}\, .
\label{singularity}
\end{equation}
The Matsubara frequency sum can be readily carried out, yielding
\begin{equation}
F[\bbox{q}^2,\omega(\bbox{p})\rightarrow 0]=
g^2_\pi(T){\tanh(m/2T)\over 4\pi m}
\Biggl\{1-{\bbox{q}^2\over 12m^2}\Bigl[1-{(m/T)\over\sinh(m/T)}\Bigr]
\Biggr\}+{\cal O}\bigl(\bbox{q}^4\bigr)\, .
\end{equation}
Using the explicit expression of $g^2_\pi(T)$ in Eq.~(\ref{g_pi})
we immediately find that the effective charge
$F[0,\omega(\bbox{p})]\le 1$, with the equality
sign only at $T=0$. The fact that $F[\bbox{q}^2,\omega(\bbox{p})]$
vanishes in the limit $T\rightarrow T_c$ merely reflects the fact that 
the cross section for a lepton to hit the pion also vanishes in the
same limit, consistent with the decoupling of the pion from its
constituents at the chiral restoration point. As another consistency
check, $F[0,\omega(\bbox{p})]$ can also be calculated from the full
pion propagator $S_\pi[\omega,\bbox{p}^2]$ through the Ward identity,
in the limit of $\bbox{p}^2\rightarrow 0$
\begin{equation}
F[0,\omega(\bbox{p})]\, =\,g_\pi^2(T)\,
{\partial\over\partial\bbox{p}^2}S_\pi^{-1}[\omega,\bbox{p}^2]
\Bigg|_{\omega=\omega(\bbox{p})}\, ,
\end{equation}
which is a consequence of the residual static gauge invariance at
finite temperature. In the above equation one must use the full pion
propagator in the limit of slow processes, i.e. taking $\omega=0$
before sending $\bbox{p}^2$ to zero as it was done in
Eq.~(\ref{bubble_p}); this
limit differs from the one used to produce Eq.~(\ref{full-bubble})
which is defined for fast processes. As we anticipated, 
the charge conservation that guarantees $F[0]=1$ at $T=0$ no longer 
holds at finite $T$, due to the well-known charge screening. 
The effective charge $F[0]$ is the pion charge plus the charge 
in its thermal cloud.
The explicit charge conservation only can be expected in a
deeply inelastic process, in which the on-shell condition is
enforced at the end of the form factor calculation. Of course, the
total charges of the test particle and the heat bath is definitely
conserved. However, in a slow scattering process, the lepton only
sees the charges localized near the test particle, but not those
charges far away at the spatial boundary.

  It would be interesting to calculate directly the thermal-cloud
charge distribution induced by a pion. However, this calculation
appears to be beyond the scope of the linear response theory. The
reason is that, according to the linear response theory, the induced 
thermal cloud is proportional (in momentum space) to the charge 
distribution of the test particle, and the temperature dependence 
appears only in the proportionality function, i.e. the retarded
correlation function. In our case, the test particle itself has
a non-trivial internal structure, which is subject to change under
the thermal environment. Therefore, there is back-reaction between
the induced thermal cloud and the structure of the pion, which is
not an external probe. This intricate entangling
of the test particle and the thermal environment necessarily
prompted us to the calculation of the on-shell form factor.

Now the radius of the pion in the heat-bath frame can be
obtained easily
\begin{equation}
\langle r^2\rangle_T=
{1\over 3m^2}\Bigl[1-{m/T\over\sinh(m/T)}\Bigr]\, .
\label{radius}
\end{equation}
The complete curve of $\langle r^2\rangle_T$ as a function of $T$
is displayed in Fig.~\ref{fig5}, which makes sense only in the
symmetry broken phase where the pion exists. At zero temperature
$\langle r^2\rangle_0 = 1/(3\mu^2)$, which can also be verified
directly from the covariant calculation at $T=0$, where
$F[q^2]=[4m^2/(-q^2)]^{1/2}\sin^{-1}[(-q^2)/(4m^2-q^2)]^{1/2}$,
with $q^2=q_0^2-\bbox{q}^2$. Near critical temperature
$\langle r^2\rangle_T\rightarrow 1/(18T_c^2)=2(\ln2)^2/(9\mu^2)$.
Although the pion's size decreases as temperature increases from
zero to the critical point by about factor of 3, the characteristic
size scale of the pion remains to be $1/\mu^2$. This is related
to the fact that the pion, once it forms, is always a tightly
bound state when $T<T_c$.

It is also interesting to point out that, in general, the radius 
$\langle r^2\rangle_T$, defined through slow processes,
can not be simply estimated by the threshold 
of the triangle graph, which is a standard practice at $T=0$.
In fact, if one did so, one would obtain the wrong estimate 
$\langle r^2\rangle_T\sim 1/m^2$, which diverges when
$T\rightarrow T_c$. The phase space integral, which produces the
factor in the brackets in Eq.~(\ref{radius}), can play an important
role at finite temperature. As can be seen clearly from
Eq.~(\ref{singularity}), the actual singularity structure of
$F[\bbox{q}^2,\omega(\bbox{p})\rightarrow 0]$ in $\bbox{q}^2$
is always controlled by the lowest threshold
$4M_1^2\equiv 4(\pi^2T^2+m^2)$, which stays of order $\mu^2$ in
the entire broken phase. A physical interpretation can be given
for Eqs.~(\ref{singularity}) and (\ref{radius}): the thermal
environment tends to wash out coherence beyond the thermal wave
length ($1/T$) in slow processes.

  At this point the physical picture is clear. At low temperature,
the pion state is very much the usual pion state with exponentially
small correction from the thermal environment. As temperature
increases the quark and antiquark become less and less likely to
bind together to form the elementary excitation. However, once they
bind together, the elementary excitation remains qualitatively similar
to its zero temperature counterpart. Above the critical temperature
the pion completely resolves into its constituents and no longer
exists. So the phase structure of the system is strongly reflected
from its spectral content in Minkowski space.

In sharp contrast, the screening Bethe-Salpeter amplitude defined
in the Euclidean space does not share this feature, since it is
insensitive to the real-time singularity structure, as we shall
demonstrate in the next section. Nonetheless, the screening
Bethe-Salpeter amplitude indeed provides a qualitative measure of
the size of the corresponding bound state, though only in the
phase where the bound state can be identified as an elementary
excitation. This is  due to the fact that the on-shell form
factor and the screening Bethe-Salpeter amplitude have thresholds
proportional to $M_1^2$.

  Can the result in Eq~(\ref{radius}) be an accident of the 2+1
dimensional Gross-Neveu model? To settle this question we need
to study the on-shell form factor in 3+1 dimensions, i.e. in the
Nambu-Jona-Lasinio model. In 3+1 dimensions, we find that the
resulting formula is not very different from the one in 2+1
dimensions, Eq.~(\ref{singularity}):
apart from some trivial factors, which are independent of the 
momentum transfer, the only modification is that the integrand of
the $\alpha$-integral is raised to the power 1/2, as it is obvious
from dimensional counting. Therefore, the lowest singularity in
$\bbox{q}^2$-plane is still given by $4M_1^2$. Although
now the form factor requires a cutoff in the Matsubara frequency
sum, the charge radius is independent of this cutoff. This implies
that the charge radius in 3+1 dimensions has
qualitatively the same behavior as in the 2+1 dimensional case.
Our result, a pion charge radius that remains finite at the critical
temperature both in the 2+1 Gross-Neveu model and in the 3+1 
Nambu-Jona-Lasinio model, is in sharp contradiction with the result 
found in Ref.~\cite{schulze}, a pion charge radius that diverges at 
the critical temperature in the Nambu-Jona-Lasinio model. Since
the result of Ref.~\cite{schulze} is not expressed through a simple
formula such as our Eqs.~(\ref{singularity}) and (\ref{radius}), but
is obtained by means of a numerical integration of an expression that
involves principal value definitions, we have not been able to pin
down the source of this discrepancy. However, it is worth mentioning
that Schulze used the temporal component of the triangle graph to
define the charge radius, which we believe is less desirable due to
the contamination from the second form factor $G$ associated with
the heat bath. It is also not clear at what exact stage he enforced
the on-shell condition, and hence it is not clear whether his form 
factor is defined for fast or slow processes. Either of these two points
could lead to very different results and, more importantly, to different
interpretations in terms of physical measurements.

\subsection{Spectral Function}
We now calculate the spectral function in the pseudoscalar channel.
We give the result for $\bbox{p}=0$, where the cut contribution near
the origin, due to the scattering of the current and thermal particles,
vanishes. Below $T_c$ we have a contribution from the continuum
($\omega> 2m$), and a pole contribution from the bound state:
\begin{equation}
\label{sf}
\rho( \omega)=\epsilon(\omega)\left[
\pi\,{N^2 g_\pi^2(T)\over g^4}\delta(\omega^2-m^2_{\pi})
+\theta( \omega^2 - 4 m^2(T))\,\rho_{\text{cont}}( \omega)
\right]\, ,
\end{equation}
where the bound state mass $m_{\pi}=0$, and the coupling
$g^2_{\pi}(T)$ is given in Eq.~(\ref{g2pi}). Above $T_c$ only 
the contribution from the continuum survives.
The continuum part of the spectral function is related to the
real and imaginary parts of the bubble graph, see
Eqs.~(\ref{psbblri}), in the following way
\begin{equation}
\label{sfcon}
\rho_{\text{cont}}( \omega) = 
\frac{\text{Im}\left[i\Pi_P(\omega)\right]}
{\left(1-(g^2/N)\text{Re}\left[i\Pi_P(\omega)\right]\right)^2 
+ \left((g^2/N)\text{Im}\left[i\Pi_P(\omega)\right]\right)^2}.
\end{equation}
We show the spectral function at different temperatures in
Fig.~\ref{fig4}. The most remarkable feature is that near the phase
transition the shape of the spectral function is very different 
from its shape at low temperature. A large peak develops just above
threshold. The peak actually diverges right at the phase transition,
though its integrated strength stays finite. At the same time, as
expected, the pole becomes weaker and the threshold goes to zero.
These qualitative features are not artifacts of the particular
model we are considering here. They actually are the characteristics
of a continuous phase transition and
critical phenomena. For example, the strong peak near the threshold
is a reflection of the fact that the corresponding susceptibility
diverges at $T_c$. These characteristics would still persist to 
a certain extent when the phase transition is weekly first order.
Furthermore, as we argued in Ref.~\cite{esr}, QCD shares all
these qualitative features. 

  The strong peak near the origin in the spectral function, when
$T$ is close to $T_c$, indicates a resonance-like excitation.
This resonance corresponds to a complex pole of the two-point
function on the second Riemann sheet of the $\omega^2$-plane,
with a very small imaginary part. Using explicit formulae in
Sec.~\ref{GN}, one can analytically continue the two-point function
onto the second Riemann sheet. Then, it is not hard to find that
the real and imaginary parts of the pole in the symmetric phase
are proportional to $t/\ln(t^{-1})$ and $-t/\ln^2(t^{-1})$,
respectively, where $t\equiv (T-T_c)/T_c\ll 1$.
As $T$ gradually increases from $T_c$
this resonance pole moves almost parallel to the real axis from
the origin to the right initially; and it eventually marches into
the first quadrant. When $T$ is outside the scaling region, the
imaginary part of this complex pole becomes so big such that
it does not makes sense anymore to call it a resonance excitation.

  It appears that the diminishing of the pion strength and the
magnification of the quark-antiquark continuum strength in the
spectral function near the critical region provides a possible
physical picture for the failure of the standard $\sigma$-model
scenario \cite{pisarski} in predicting the nature of the
chiral-symmetry-restoration phase transition \cite{kocic} in
the Gross-Neveu model in 2+1 dimensions. To substantiate this
statement a detailed investigation is necessary. The reason is
that the $\sigma$ field in the relevant $\sigma$-model does not
necessarily correspond to any real-time bound state pole, but
more likely to the resonance pole mentioned above.

\subsection{Sum Rule}
To quantitatively characterize the weakening of the pole and
the growing of the peak above threshold we derive an exact
sum rule for the temperature dependent part of the
zeroth moment of the spectral function.

To derive the sum rule we need to find out the asymptotic behaviors
of the two-point function in the deep Euclidean region and of the
spectral function in the large-$\omega^2$ limit. Using the expression
in Eq.~(\ref{psbbl}) it is easy to obtain the pseudoscalar bubble
graph in the Euclidean region
\begin{equation}
i\Pi_P(iQ,\bbox{p}\rightarrow 0)={N\over g^2}+{N\over\pi}
\biggl\{ R(T)-{Q\over 2}\tan^{-1}{Q\over 2m}
+{Q^2\over 4}\int_m^\infty dE\, {2\over 1+e^{E/T}}\,
{1\over E^2+Q^2/4}\biggr\}\, ,
\end{equation}
and its asymptotic form at large-$Q^2$
\begin{equation}
{1\over N}i\Pi_P(iQ,\bbox{p}\rightarrow 0)\sim -{Q\over 4}
+\biggl({1\over g^2}+{\mu\over\pi}\biggr)
-{4\over 3\pi}\,{m^3+3\langle\!\langle E^2\rangle\!\rangle\over Q^2}
+{\cal O}(Q^{-4})\, ,
\end{equation}
where $\langle\!\langle\cdots\rangle\!\rangle$ stands for the 
thermal average
\begin{equation}
\langle\!\langle A\rangle\!\rangle\equiv\int_m^\infty
dE \,{2\over 1+e^{E/T}}\, A(E) \, .\nonumber
\end{equation}
For example, $\langle\!\langle E^2\rangle\!\rangle=3\zeta(3)T^3$,
in the symmetric phase where $m=0$. The corresponding two-point
function, defined in Eq.~(\ref{twopoint}), is
\begin{eqnarray}
\langle J_5 J_5\rangle_T(iQ,0)
&\sim&-{N\over g^2}\biggl\{1-{4\over g^2 Q}
\biggl[1+{4\mu\over\pi Q}+{16\mu^2\over\pi^2 Q^2}+
{64\mu^3-16\pi^2(m^3+3\langle\!\langle E^2\rangle\!\rangle)/3
\over \pi^3 Q^3}\biggr]\biggr\} \nonumber \\
&+&{\cal O}(Q^{-5})\, .
\label{J5_asym}
\end{eqnarray}

  On the other hand, the spectral function has the following
asymptotic form at large $\omega^2$, as can be derived from
Eqs.~(\ref{psbblri}) and (\ref{sfcon}),
\begin{equation}
\rho(\omega)\sim {4N\over g^4\omega}
\biggl\{1-16\biggl({\mu-R(T)\over\pi\omega}\biggr)^2\biggr\}
+{\cal O}(\omega^{-5})\, .
\label{rho_asym}
\end{equation}
To study the temperature dependence we make the subtraction,
following Ref.~\cite{esr},
\begin{equation}
\Delta\langle J_5 J_5\rangle(iQ,0)\equiv
\langle J_5 J_5\rangle_T(iQ,0)-\langle J_5 J_5\rangle_{T'}(iQ,0)
=\int_0^\infty d\omega^2
{\Delta\rho(\omega)\over \omega^2+Q^2}\, ,
\end{equation}
with $\Delta\rho(\omega)\equiv\rho_T(\omega)-\rho_{T'}(\omega)$.
Since the leading term in $\rho_T(\omega)$ is independent of $T$,
as seen from Eq.~(\ref{rho_asym}), we immediately have
$\Delta\rho\sim \omega^{-3}$. This implies that the zeroth moment
of $\Delta\rho$ exists. 
Using Eq.~(\ref{J5_asym}), and hence
\begin{equation}
\Delta\langle J_5 J_5\rangle(iQ,0)\sim{64N\over 3\pi g^4}
{\Delta\bigl(m^3+3\langle\!\langle E^2\rangle\!\rangle\bigr)
\over Q^4}+{\cal O}(Q^{-5})\, ,
\end{equation}
we arrive at the following expression appropriate to derive
the sum rule
\begin{equation}
{64N\over 3\pi g^4}
{\Delta\bigl(m^3+3\langle\!\langle E^2\rangle\!\rangle\bigr)
\over Q^2}=\int_0^\infty d\omega^2 {\Delta\rho\over 1+\omega^2/Q^2}
+{\cal O}(Q^{-3})\, ,
\end{equation}
where $\Delta\bigl(m^3+3\langle\!\langle E^2\rangle\!\rangle\bigr)
\equiv\bigl(m^3+3\langle\!\langle E^2\rangle\!\rangle\bigr)_T
-\bigl(m^3+3\langle\!\langle E^2\rangle\!\rangle\bigr)_{T'}$.
Taking the limit $Q^2\rightarrow\infty$ we get the desired sum rule
\begin{equation}
\int_0^\infty d\omega^2 \, \Delta\rho=0\, ,
\label{sum_rule}
\end{equation}
Interchanging the order of the dispersion integral and the
$Q^2\rightarrow\infty$ limit is allowed since the zeroth
moment is finite.

  We have proved that the very same sum rule also holds in
the 1+1 dimensional Gross-Neveu model and QCD~\cite{esr}, though
we have used a totally different derivation in those cases and
the corresponding sum rules would only converge logarithmically.
It appears that the temperature dependence of the zeroth moment 
of the spectral function often shows qualitatively similar behaviors, 
no matter what is the underlying microscopic physics. 

Another interesting
point, which was especially clear in the derivation we used for
the 1+1 dimensional Gross-Neveu model and QCD~\cite{esr}, is that
the zeroth moment of the spectral function could be related, in
asymptotically free theories, to expectation values of 
appropriate operators via the operator product expansion. That
result cannot be immediately applied to the 2+1 Gross-Neveu model,
since this model is not asymptotically free and possesses a
nontrivial ultra-violet fixed point. Therefore, it would be
extremely interesting to investigate whether a formalism similar
to the operator product expansion~\cite{rge} can be extended 
to this case.

\section{Screening phenomena}
\label{screen}
As discussed in section~\ref{realt}, screening phenomena are
associated with responses to time-independent external
perturbations. Therefore, we only need to consider static Green's
functions and purely spatial correlation functions. Static Green's
functions are essentially Euclidean and hence, in contrast to the
time-dependent responses we have discussed in the preceding section,
well-suited for the lattice formulation. In this section we study
the static two-point and three-point functions that relate to
screening masses and screening Bethe-Salpeter amplitudes,
respectively, since these quantities are studied on the 
lattice~\cite{mscreen,latbs}.
\subsection{Screening mass} 
  The pseudoscalar screening mass $\tilde{m}$ is the lowest
solution to the equation
\begin{equation}
0=1-{g^2\over N} \Pi_P(i\omega\rightarrow 0,i\bbox{p})=
-\frac{g^2}{\pi}\left\{R(T)+2T\sum_{n=-\infty}^\infty
\sqrt{\bbox{p}^2\over 4M_n^2-\bbox{p}^2}
\,\tan^{-1}\sqrt{\bbox{p}^2\over 4M_n^2-\bbox{p}^2}\right\}\, .
\label{screenp}
\end{equation}
When $R(T)=0$ ($T<T_c$), the solution is obvious:
$\bbox{p}^2=0=\tilde{m}_\pi^2$. In the high temperature phase 
$R(T)\neq 0$, the solution can only be obtained numerically. 

Contrary to the naive expectation that higher than lowest
Matsubara modes decouple at high $T$, we point out that the 
frequency sum in Eq.~(\ref{screenp}) cannot be truncated without
introducing an error in the screening mass that is not of order $1/T$.
This fact suggests that dimensional reduction does not take place in
this model~\cite{drl}. Details on this point will be presented later.
 
  We can define a coupling constant for the screening state
analogous to the one defined for the real-time bound state case
\begin{equation}
\tilde{g}_\pi^2(T)=\left({\partial\over\partial\bbox{p}^2}
\Pi_P(i\omega\rightarrow 0,\bbox{p})
\right)^{-1}_{\bbox{p}^2=-\tilde{m}_\pi^2}\, .
\end{equation}
Then, near the screening-mass pole, the static pseudoscalar
propagator behaves like
$\tilde{g}_\pi^2(T)/(\bbox{p}^2+\tilde{m}_\pi^2)$.
In Fig.~\ref{fig2} $\tilde{g}^2_\pi(T)$ is plotted in dashed line.
Even though $\tilde{g}^2_\pi$ agrees with $g_\pi^2$ at $T=0$, due
to the Euclidean invariance, it becomes drastically different
from $g_\pi^2$ as temperature increases. In particular, 
$\tilde{g}_\pi^2$ does not vanish in the symmetric phase, 
but rather grows linearly with $T$ in the high-$T$ limit.
In the broken phase $\tilde{g}_\pi^2(T)$ has a particularly
simple form: $\tilde{g}_\pi^2(T)=4\pi m/[N\,\tanh(m/2T)]$.

  As in the real-time case, an analogous ``decay constant'' for
the ``on-shell'' screening pion can be defined. Again it is easy
to verify
\begin{equation}
\tilde{f}_\pi(T)=\tilde{g}_\pi(T)\,{N\over 2\pi}\,
\tanh\left({m(T)\over 2T}\right)\, ,
\label{fpiim}
\end{equation}
which also vanishes in the symmetric phase. It is interesting
to notice that the ``Goldberger-Treiman'' relation holds exactly
in the screening case
\begin{equation}
\tilde{g}_\pi(T)\, \tilde{f}_\pi(T)=2m(T)\, .
\end{equation}
We believe that this result is not accidental, because the
screening state in the static limit can be regarded as a bound state 
in some 1+1 dimensional theory at zero temperature with the same
symmetry as the original theory. This Euclidean ``Goldberger-Treiman''
relation is in agreement with the results of Ref.~\cite{shen}.
 
  In order to explicitly see the manifestation of the chiral
symmetry also in the screening masses, it is instructive to
calculate the scalar screening mass. The scalar bubble graph (in 
the static limit) is easily related to the pseudoscalar counterpart
\begin{equation}
\Pi_S(i\omega\rightarrow 0,\bbox{p})=
\Pi_P(i\omega\rightarrow 0,\bbox{p})
-{NT\over\pi}\sum_{n=-\infty}^\infty {4m^2\over\bbox{p}^2}
\sqrt{\bbox{p}^2\over 4M_n^2+\bbox{p}^2}
\,\ln\left[{\sqrt{4M_n^2+\bbox{p}^2}+\sqrt{\bbox{p}^2}\over
\sqrt{4M_n^2+\bbox{p}^2}-\sqrt{\bbox{p}^2}}\right]\, .
\end{equation}
Similarly to the pseudoscalar screening mass, the scalar screening
mass is found by solving the equation
\begin{eqnarray}
0&=&1-{g^2\over N} \Pi_S(i\omega\rightarrow 0,i\bbox{p})\nonumber\\
&=&-\frac{g^2}{\pi}\left\{R(T)+2T
\sum_{n=-\infty}^\infty {\bbox{p}^2-4m^2\over\bbox{p}^2}
\sqrt{\bbox{p}^2\over 4M_n^2-\bbox{p}^2}
\,\tan^{-1}\sqrt{\bbox{p}^2\over 4M_n^2-\bbox{p}^2}\right\}\, .
\label{screens}
\end{eqnarray}
When $T<T_c$, $R(T)=0$  and the solution is again easily obtained
$\bbox{p}^2=4m^2=\tilde{m}_{\sigma}^2$ 
(notice that $\bbox{p}^2=0$ is not a solution in this 
channel). For $T>T_c$ the dynamical quark mass $m^2(T)$ vanishes,
and the scalar screening mass equation becomes identical to the
one for the pseudoscalar screening mass.
 
  In Fig.~\ref{fig6} we display numerical results for the
screening masses in both channels, pseudoscalar and scalar. In the
low temperature phase, chiral symmetry is broken, and the pion is
the relevant Goldstone boson: this fact is also reflected in the
screening mass.  In the high temperature phase, chiral
symmetry is restored, and we find the expected parity doubling,
i.e. the degeneracy of the screening masses in the parity mirrored 
channels. 

   We already know that the screening state is very different from
the real-time bound state. One might, however, naturally ask whether
the screening state be somehow related to the resonance excitation
near $T_c$. In particular, one might want to introduce some non-trivial
external momentum dependence in the screening state and hope that, by
making this external spatial momentum complex, the screening pole might
be continued into the resonance pole. While this option is in principle
possible, we think that a more relevant question is whether the knowledge
of the static screening state as a function of temperature is sufficient
by itself, i.e. when we do not have complete knowledge of the analytic
structure of the two-point function, to provide enough direct information
on the resonance state. We think that the answer to this latter question
is unlikely to be affirmative. The reason is that the functional 
dependences of the two-point function on frequency and spatial momemtum 
at finite $T$ are genuinely different, due to the explicit lack of Lorentz 
invariance. Therefore, it is not possible to probe, in general, the full 
information carried by the frequency variable by varying a spatial 
component of the momentum.
In addition, the following three points also support this general view.
First, the screening pole is purely imaginary for any $T$, whereas the
resonance pole is complex in general. Second, the screening state is well
defined for any $T$, whereas the resonance is only identifiable near the
critical region. The third point is more specific to the model we have
consider: when $T\gtrsim T_c$, the screening mass squared is proportional
to $t$, with $t=(T-T_c)/T_c\ll 1$, whereas the resonance pole is 
proportional to $t$ with logarithmic corrections (see section IV.D).

\subsection{Screening Bethe-Salpeter amplitude}
Another observable that is calculated on the lattice at finite 
temperature~\cite{latbs} is the screening Bethe-Salpeter amplitude
(SBSA). Our definition of this amplitude in the pseudoscalar channel
is
\begin{equation}
\phi(x,y)\equiv\int d\tau\,dz\,\langle \bar{\psi}(\tau,x,z)\gamma_5
\psi(\tau,x,z+y)\,\bar{\psi}(0,0,0)\gamma_5\psi(0,0,0)\rangle\, .
\end{equation}
This definition differs from the usual lattice definition by an
extra integration over $\tau$. We perform this extra integration for
computational convenience, since it allows us to consider only the
zero frequency mode. In terms of physical content of the SBSA, this
difference should be irrelevant. To the leading order in $1/N$, the 
Feynman graphs contributing to the SBSA are
those shown in Fig.~\ref{fig7}. Then the
Fourier transform of $\phi(\tau,y)$ has the following form
\begin{equation}
\tilde{\phi}(p_1,p_2)={\tilde{\phi}_1(p_1,p_2)\over
1-g^2 \Pi_P(i\omega\rightarrow 0,\bbox{p}=(p_1,0))/N}\, ,
\end{equation}
where $\tilde{\phi}_1(p_1,p_2)$ is the off-shell screening BS
amplitude
\begin{equation}
\tilde{\phi}_1(p_1,p_2)
={iNT}\sum_n \int {d^2\bbox{k}\over(2\pi)^2}
\text{Tr}\left[i\gamma_5\, {i\over k\cdot\gamma-m}\, i\gamma_5\,
{i\over k\cdot\gamma-p\cdot\gamma-m}
\right]_{p=(i\omega\rightarrow 0,p_1,p_2)}\, .
\end{equation}
An explicit calculation yields
\begin{equation}
\tilde{\phi}_1(p_1,p_2)={2NT}\sum_{n=-\infty}^\infty
{\sqrt{M_n^2+p_2^2}\over M_n^2+p_2^2+p_1^2/4}\, .
\end{equation}
Therefore, the complete SBSA is
\begin{equation}
\phi(x,y)=\int{dp_2\over 2\pi} \, e^{-ip_2 y}
\int{dp_1\over 2\pi}\, e^{-ip_1 x}{\tilde{\phi}_1(p_1,p_2)
\over 1-g^2 \Pi_P(p=(i\omega\rightarrow 0,p_1,0))/N}\, .
\end{equation}
 
Typically, we are interested in the dependence of the SBSA on $y$
for large $x$, so that we project on the lowest ``screening state''.
In the limit $x\rightarrow\infty$ the $p_1$--integral picks out
the lowest singularity, $p_1^2=\tilde{m}_\pi^2<4M_n^2$, and the
SBSA becomes
\begin{equation}
\phi(x,y)\bigg|_{x\rightarrow\infty}
\longrightarrow{\tilde{g}_\pi^2 N\over 2\tilde{m}_\pi}
\, e^{-\tilde{m}_\pi x} \sum_{n=-\infty}^\infty 
{2T}\int{dp_2\over 2\pi} \, e^{-ip_2 y}
{\sqrt{M_n^2+p_2^2}\over M_n^2-\tilde{m}_\pi^2/4+p_2^2}\, .
\label{sbsa}
\end{equation}
The scale that controls the size of the screening
wave function is clearly $\sqrt{M_1^2-\tilde{m}_\pi^2/4}$, as one
can show by deforming the integration contour to the imaginary axis.
Since $M_1^2=(\pi T)^2 + m^2$, and $m_{\pi}=0$ at low temperature,
in the low $T$ limit this scale is $m\propto \mu$. At high
temperature $m=0$, and $m_{\pi}\propto T$, therefore in this limit
the scale is set by $\sqrt{\pi^2 T^2-\tilde{m}_\pi^2/4}\sim T$.
 
As we have argued in the preceding section, a better measure of
the spatial distribution of the quark and antiquark inside a
meson is provided by its on-shell form factor. 
The SBSA qualitatively measures the size of the elementary 
excitation, or bound state, in the entire broken phase.
However, the SBSA above the phase transition
does not yield any information about the nature of the 
corresponding (nonexistent) real-time bound state.

  We also calculate the SBSA in the scalar channel using exactly
the same procedure we have used in the pseudoscalar channel, and
we find very similar results. The relevant mass scale is now given
by $\sqrt{M_1^2-\tilde{m}_\sigma^2/4}$, which has the same high
temperature limit as the scale for the pseudoscalar channel. In the
broken phase ($T<T_c$), since $\tilde{m}^2_{\sigma}=2m$, and
$M_1^2=(\pi T)^2 + m^2$, $\sqrt{M_1^2-\tilde{m}_\sigma^2/4}=\pi T$.
This last result is consistent with the fact that there is no
binding energy in the scalar channel, and the only screening is
due to the thermal mass. 

\subsection{Real-time vs. screening phenomena}
  Now we have all the ingredients to make an explicit comparison
between the real-time bound state and the screening state.

  {\em Mass}: When $T<T_c$ (the chiral symmetry is spontaneously
broken), both the real-time and the screening pion masses vanish,
$m_\pi(T)=\tilde{m}_\pi(T)=0$, while the scalar mass
$\tilde{m}_\sigma(T)=2m(T)$. However, when $T>T_c$ (chirally
symmetric phase), the real-time pion decouples and the spectral 
function has contributions only from the quark-antiquark continuum, 
whereas the screening pion mass $\tilde{m}_{\pi}(T)$ becomes
degenerate with the screening scalar mass $\tilde{m}_\sigma(T)$
and both of them asymptotically grow linearly with $T$
(see Fig.~\ref{fig6}). 

 {\em Coupling}: The real-time coupling of the pion to the 
quark-antiquark state, $g_\pi^2(T)$, and the corresponding
screening coupling, $\tilde{g}^2_\pi(T)$, have very different
behaviors as functions of temperature, as it is clearly shown
in Fig.~\ref{fig2}, and they only coincide at $T=0$.
In particular, the screening coupling $\tilde{g}^2_\pi(T)$ does
not provide any signal for the fact that the real-time excitation
decouples when $T\geq T_c$; in fact $\tilde{g}^2_\pi(T)$
does not show any distinctive feature near the critical region.

 {\em Decay constant}: Both $f_\pi(T)$ and $\tilde{f}_\pi(T)$
start at the same value at $T=0$ and gradually decrease to zero
at $T=T_c$. The vanishing of $f_\pi(T)$ and $\tilde{f}_\pi(T)$
when $T\ge T_c$ is due to the fact that the axial current decouples
from the pseudoscalar current when the chiral symmetry is restored.
This constraint forces a similar behavior for both $f_\pi(T)$ and 
$\tilde{f}_\pi(T)$, even if their numerical values between $T=0$
and $T=T_c$ are different. Above the phase transition, 
$f_\pi$ loses its meaning, and $\tilde{f}_\pi(T)$ remains zero.

 {\em Size}: As seen from Fig.~\ref{fig5}, the sizes
defined through the on-shell form factor in the elastic limit,
$\langle r^2\rangle_T$
and the screening Bethe-Salpeter amplitude, $1/M_1^2$, have
qualitatively similar behaviors in the broken phase. The reason
for this similarity is that they share the same singularity structure 
in low spatial momentum transfer. Nevertheless, we find again that,
while $\langle r^2\rangle_T$ loses its meaning above $T_c$, the size
defined through the screening Bethe-Salpeter amplitude does not
give us any signal of this disappearance of the real-time state.

  We believe that most of the above-mentioned qualitative features 
will also be present in 3+1 dimensions, although some of the details 
presented here are certainly specific to this model. Therefore, we
can safely conclude that the screening observables, in general, do
not necessarily reflect the behavior of the corresponding real-time 
observables. These two kinds of observables are often qualitatively
different.

\subsection{High-$T$ limit and dimensional reduction}

  The present model study can also offer us very interesting
information about the possibility of dimensional reduction
for screening Green functions at high temperature.

  In the symmetric phase, equations~(\ref{screenp}) and
(\ref{screens}) coincide and can be rewritten as
\begin{equation}
{\mu\over T_c}-{\mu\over T}=4\sum_{n=1}^\infty
{x\over\sqrt{4(2n-1)^2-x^2}}
\tan^{-1}{x\over\sqrt{4(2n-1)^2-x^2}}\, ,
\end{equation}
where $x^2\equiv \bbox{p}^2/(\pi T)^2$. In the high-$T$ limit
we can solve the above equation iteratively by setting 
$x=\sum_{i=0}^\infty c_i\,(\mu/\pi T)^i$. In particular,
the first two coefficients can be easily found by solving
\begin{equation}
{\mu\over T_c}=4\sum_{n=1}^\infty
{c_0\over\sqrt{4(2n-1)^2-c_0^2}}
\tan^{-1}{c_0\over\sqrt{4(2n-1)^2-c_0^2}}\, ,
\label{c0}
\end{equation}
and
\begin{equation}
{\mu\over c_1}=-{4\over \pi}
\sum_{n=1}^\infty \left\{ {c_0\over 4(2n-1)^2-c_0^2}
+{4(2n-1)^2\over [4(2n-1)^2-c_0^2]^{3/2}}
\tan^{-1}{c_0\over\sqrt{4(2n-1)^2-c_0^2}} \right\} \, .
\label{c1}
\end{equation}
Therefore, the asymptotic form for the screening
masses in the high-$T$ limit is,
\begin{equation}
\tilde{m}_{\pi,\sigma}=c_0\,\pi T+c_1\,\mu+{\cal O}(\mu/T)\, .
\label{msasymp}
\end{equation}
Numerically, $c_0\approx 0.982$ and $c_1\approx -0.947$.
Eq.~(\ref{msasymp}) is plotted in dashed line in Fig.~\ref{fig6}.
A salient feature of this asymptotic behavior is that 
the high temperature limit of the screening mass is $c_0\,\pi T$,
where the numerical value of $c_0$ has contributions from all the
Matsubara frequency. If dimensional reduction took place in the 
high-$T$ limit, we would find that only the lowest (or at most a
finite number) modes would contribute to $c_0$. In fact, any
truncation in the frequency sum in Eq.~(\ref{screenp}), and then
in Eq.~(\ref{c0}), would give a different $c_0$: the resulting
error in the screening mass would not be suppressed in the
high-$T$ limit.

  The reason for the failure of the dimensional reduction picture
is that the 2+1 dimensional Gross-Neveu model lacks the necessary
scale hierarchy: there is no elementary bosonic particle whose
zero mode could dominate over other modes of order $T$, and the
coupling constant cannot provide a second lower scale compared to
$T$, as it happens in asymptotically free theory~\cite{drl},
since this model has a nontrivial ultraviolet fixed point. 
In fact, the high-momentum behavior of the dimensionless coupling 
$G(\kappa)\equiv\kappa g^2(\kappa)$ has the form $G^*/(1-\mu/\kappa)$,
with $G^*=\pi$ (the value of $G^*$ is scheme dependent).
This nontrivial fixed point implies that this dimensionless coupling 
``runs'', in the high temperature limit, to a finite value
(since $\kappa\sim T$~\cite{drl}). Therefore the effective
interaction strength at high $T$ between all the Matsubara 
modes ($\sim T G(T)$, once the fields have been appropriately 
re-scaled) is strong and its strength in units of $T$ becomes
independent of temperature ($\approx T G^*$).

In this regard the Gross-Neveu model in 1+1 dimensions is very
different from the one in 2+1 dimensions. The 1+1 dimensional model
is asymptotically free and $g^2(T)$ runs to zero logarithmically,
providing the scale hierarchy between $T$ and
$g^2(T)T$~\cite{drl}. In this model dimensional reduction takes
place and screening quantities are well reproduced by the lowest
modes that are weakly interacting: the leading contribution to the 
screening mass is given by the free theory, $c_0=2$, the first 
correction to this non-interacting behavior is of the order $g^2(T)$
and it is correctly reproduced by the reduced theory. Since $g^2(T)$
runs to zero the reduced theory becomes exact asymptotically.

  Finally, we should point out that the concept of dimensional 
reduction that we have discussed in this subsection and in
Refs.~\cite{drl,drqcd}, is different from the concept of
dimensional reduction employed for predicting the nature of
phase transitions at finite $T$ in the $\sigma$-model 
scenario~\cite{pisarski}. Our criterion for dimensional reduction
is more stringent and applies to a different temperature regime.
We require that the static Green's functions of the original
theory at high-$T$ limit be matched by the corresponding
Green's functions in the reduced theory to a given accuracy,
whereas in the $\sigma$-model scenario one requires that the
reduced theory be able to match the thermodynamical singularities
near the critical region and hence give the same critical exponents.
Therefore, the success/failure of one of the two dimensional
reduction pictures does not directly implies the success/failure
of the other.
\section{Conclusion}
\label{concl}
  We have shown in a specific example, the Gross-Neveu model in 2+1
dimensions, that the direct connection between the real-time bound
state and the corresponding screening state is lost at finite
temperature in general, especially near and above phase transition.

  In particular, the real-time pion disappears after the phase
transition, as it is signaled by the vanishing of its bounding
energy and of the coupling to its constituents $g_\pi^2(T)$. 
In spite of this, the corresponding screening state is still strongly 
``bound'' in the high temperature limit, and the screening coupling 
$\tilde{g}_\pi^2(T)$ grows linearly with $T$. The screening mass
approaches the asymptotic value of $0.982\pi T$, which is much smaller
than the sum of the masses of two noninteracting fermions, $2\pi T$.

  Moreover, the screening Bethe-Salpeter amplitude yields sizes of 
order $1/T$ both in the scalar and pseudoscalar channels at high $T$,
when the real-time bound state no longer exists. Therefore, no
relevant information about the nature of the real-time pion can be
inferred from this screening amplitude. On the contrary, we can give
a natural definition of the size of the real-time pion (below the
phase transition) by means of the on-shell form factor in the elastic
limit, since the on-shell
form factor has direct information of the relevant coupling
$g_\pi^2(T)$. Furthermore, we show that the associated size cannot
be estimated solely from simple threshold considerations, reflecting
the fact that the thermal environment tends to wash out coherence
beyond thermal wavelength in slow (compared to the equilibration
time of the thermal bath) processes.

  We have also computed the exact pseudoscalar spectral function of
this model. The most important features of this spectral function are
the diminishing of the strength of the pion pole and the magnification
of the strength of the quark-antiquark continuum near the phase
transition. In addition, we have derived an exact sum rule for this
spectral function: the integrated strength of the spectral function
is independent of $T$. 

  Many of the features of the 2+1 dimensional Gross-Neveu model that
we have listed above are dominated by the chiral symmetry and should
have their analogues in the real world. For instance, very similar
results were found in the 1+1 dimensional Gross-Neveu model~\cite{esr},
and in the Nambu-Jona-Lasinio model~\cite{hatsuda} in 3+1 dimensions.
On the other hand, certain specific aspects certainly depend on
the details of the theory. For example, we find that the 2+1
dimensional Gross-Neveu model does not undergo dimensional reduction
in the high-$T$ limit. But the possibility of using a lower
dimensional theory to describe static correlations of the original
theory when $T$ is large is strictly tied to the short distance
behavior, which is very different in this model from the one in
QCD~\cite{drqcd}.

  Finally, we also find that it is instructive to classify physical
processes according to their characteristic time scales relative
to the time scale required by the thermal environment to respond to
an external perturbation. We have explicitly illustrated, in the
cases of the Goldberger-Treiman relation and of the effective charge
of the pion, that conservation laws due to internal symmetries are
manifest only in fast processes but not in slow ones.

  It is our hope that all the lessons learned from this exactly
soluble model study will provide some useful insights to QCD.

\acknowledgments
It is our pleasure to thank A.~Kocic and J.~Kogut for their valuable
comments on the manuscript, for bringing several relevant references
to our attention and for very instructive discussions on topics related
to this work. We would also like to thank T.~Hatsuda and A.~Ukawa 
for several useful discussions. One of us (ML) gratefully
thanks the Department of Energy's Institute for Nuclear Theory
at the University of Washington, where this work was initiated,
for their hospitality and partial support during
the workshop on ``Phenomenology and Lattice QCD''.
This work was supported in part by funds provided by the U.S.
Department of Energy (DOE) under contract number DE-FG06-88ER40427
and cooperative agreement DE-FC02-94ER40818.

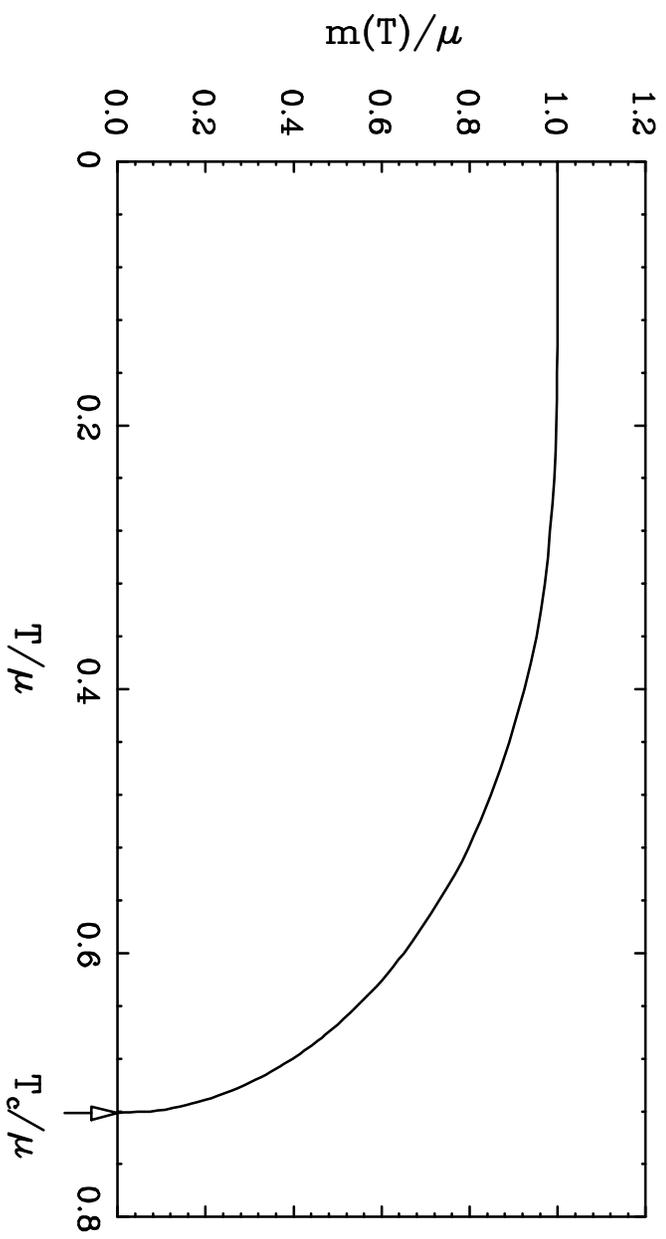
\begin{figure}
\caption[mass]{The dynamical mass as a function of temperature.
The mass and $T$ are in units of $\mu$, which is 
the dynamical mass at $T=0$.}
\label{fig1}
\end{figure}
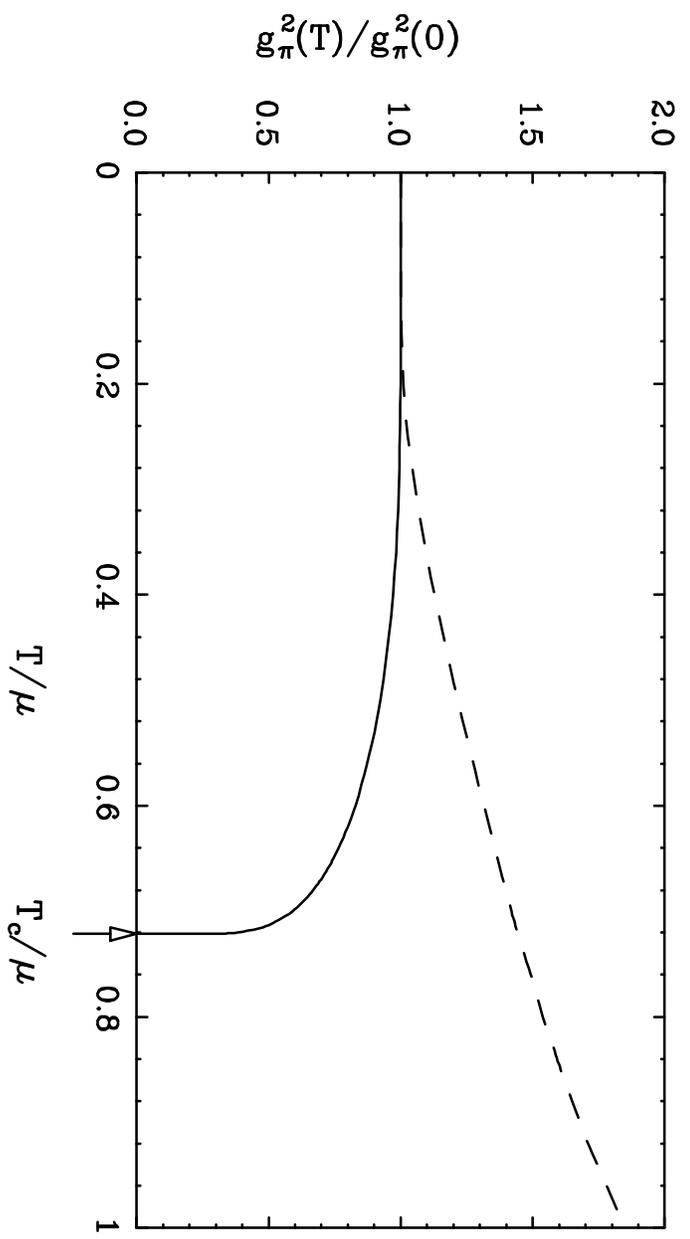
\begin{figure}
\caption[g2pif]{The bound state coupling constant, $g_\pi^2(T)$,
as a function of temperature (solid line). The dashed line is the
coupling constant for the screening state $\tilde{g}_\pi^2(T)$.
Both couplings are in units of their zero temperature value: 
$4\pi\mu/N$. }

\label{fig2}
\end{figure}
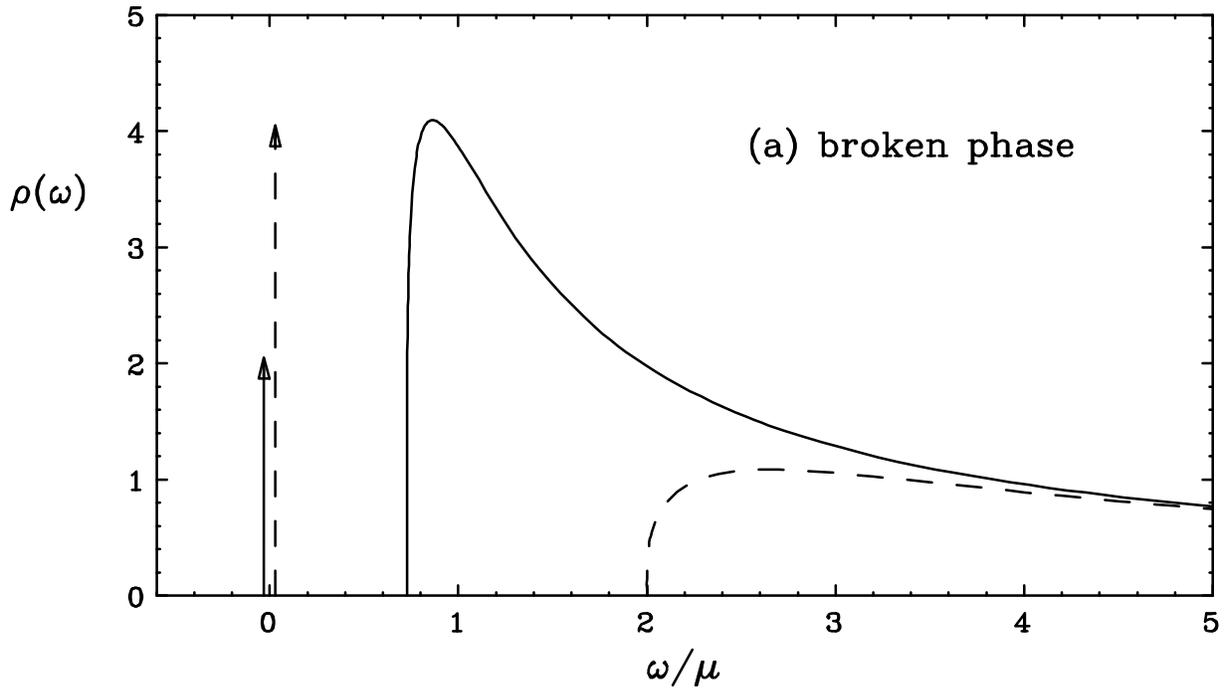
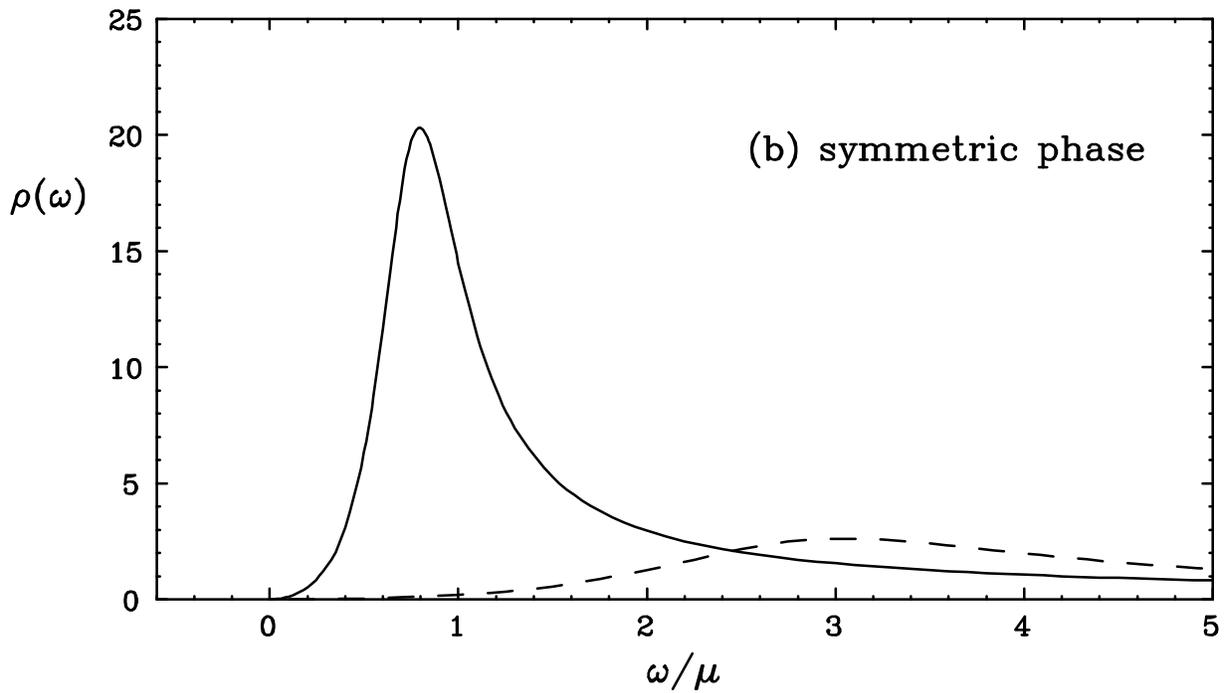
\begin{figure}
\caption[sf]{The spectral function in the pseudoscalar channel at
$\bbox{k}=0$ as a function of $\omega/\mu$: (a) in the broken phase
at $T=0$ (dashed line) and $T=0.9524\,T_c$ (solid line); and (b) in
the symmetric phase at $T=1.25\,T_c$ (solid line) and $T=2.5\,T_c$
(dashed line). The arrows in (a) denote the pion poles at $\omega=0$,
with a slight displacement for visual clarity. The height of the
arrows indicates the relative strengths of the poles.}
\label{fig3}
\end{figure}
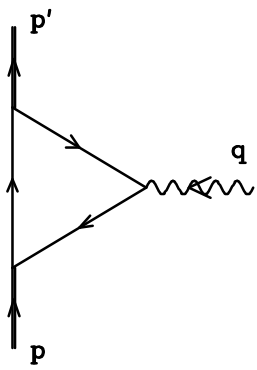
\begin{figure}
\caption[tpf]{Feynman graph for the three-point function, whose
on-shell value gives the form factor. The thick external lines
are the pion lines and the wiggle line is the photon line.}
\label{fig4}
\end{figure}
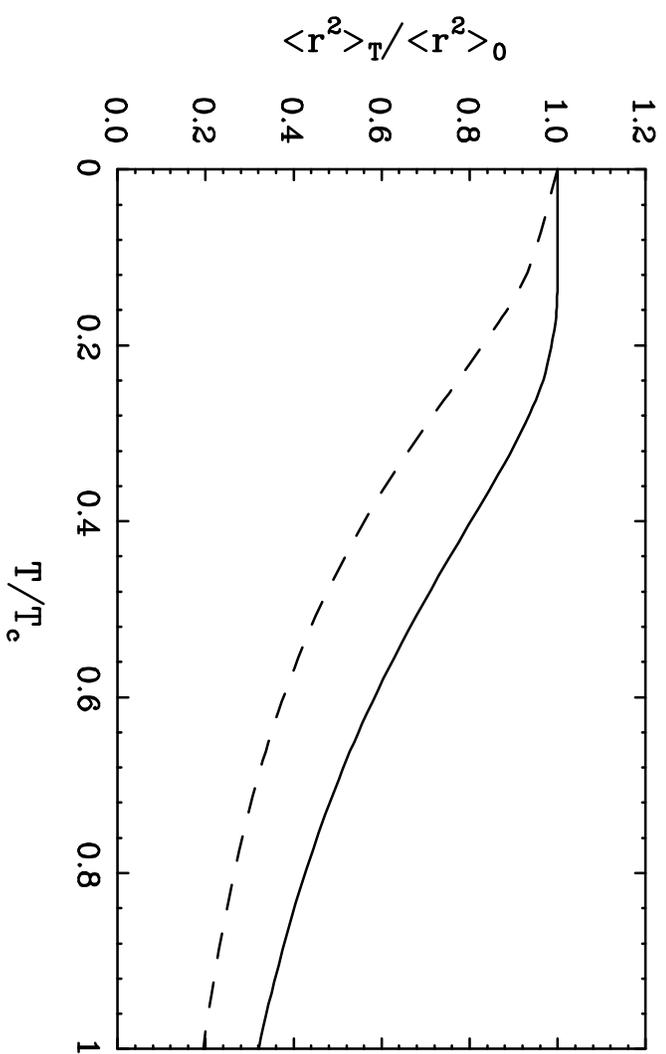
\begin{figure}
\caption[r^2]{On-shell charge radius, normalized
to its value at $T=0$, $\langle r^2\rangle_T/\langle r^2\rangle_0$,
as a function of temperature in the symmetry broken phase. 
For comparison, the ``size'' estimated from the screening
Bethe-Salpeter amplitude, $1/M_1^2$, is also plotted in dashed line.}
\label{fig5}
\end{figure}
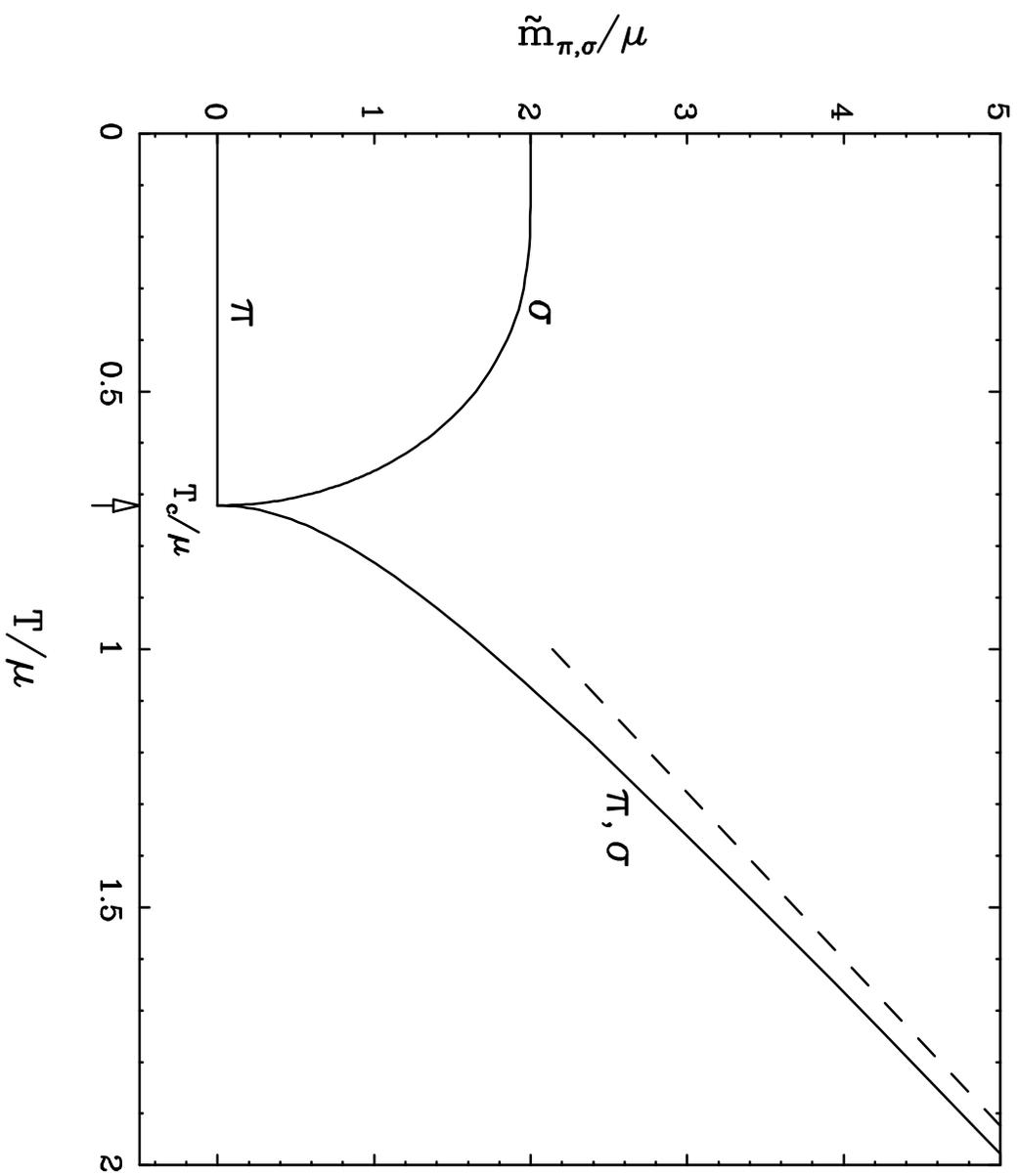
\begin{figure}
\caption[sm]{Screening masses in the scalar ($\sigma$) and 
pseudoscalar ($\pi$) channels in units of $\mu$.
It is also plotted in dashed line the asymptotic formula at high-$T$
limit, $\tilde{m}_{\sigma}=\tilde{m}_{\pi}=0.982\,\pi T-0.947\mu$.}
\label{fig6}
\end{figure}
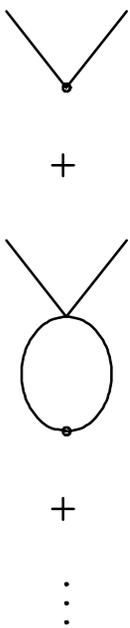
\begin{figure}
\caption[SBSAfg]{Feynman graphs for the screening Bethe-Salpeter
amplitude.}
\label{fig7}
\end{figure}
\end{document}